\documentclass[a4paper,fleqn,usenatbib]{mnras}

\usepackage{newtxtext,newtxmath}

\usepackage[T1]{fontenc}
\usepackage{ae,aecompl}

\usepackage{graphicx}
\usepackage{lscape,psfig}

\newcommand{\pedix}[2]{\ensuremath{#1_{\,\mbox{\scriptsize #2}}}}
\newcommand{\apix}[2]{\ensuremath{#1^{\,\mbox{\scriptsize #2}}}}
\newcommand{\pedap}[3]{\ensuremath{#1_{\,\mbox{\scriptsize #2}}^{\,\mbox{\scriptsize #3}}}}
\newcommand{\pedSM}[2]{\ensuremath{#1_{\,\mbox{\tiny #2}}}}

\newcommand{\errUD}[2]{\ensuremath{^{+#1}_{-#2}}}

\newcommand{\eg}{{e.g.}}

\newcommand{\mydag}{\ensuremath{\mbox{\dag}}}
\newcommand{\myddag}{\ensuremath{\mbox{\ddag}}}
\newcommand{\nustar}{\emph{NuSTAR}}
\newcommand{\spitzer}{{\emph{Spitzer}}}
\newcommand{\swift}{{\emph{Swift}}}
\newcommand{\xmm}{{XMM-\emph{Newton}}}
\newcommand{\suzaku}{\emph{Suzaku}}
\newcommand{\flux}{\ensuremath{\mbox{ergs~cm}^{-2}\mbox{~s}^{-1}}}
\newcommand{\nh}{\ensuremath{\mbox{cm}^{-2}}}
\newcommand{\kms}{\ensuremath{\mbox{km~s}^{-1}}}
\newcommand{\lum}{\ensuremath{\mbox{ergs~s}^{-1}}}
\newcommand{\arcdeg}{\ensuremath{^{\circ}}}
\newcommand{\ionpar}{\ensuremath{\mbox{ergs~cm~s}^{-1}}}

\newcommand{\normGAUSS}{\ensuremath{\mbox{photons~cm}^{-2}\mbox{~s}^{-1}}}
\newcommand{\normPL}{\ensuremath{\mbox{photons~keV}^{-1}\mbox{~cm}^{-2}\mbox{~s}^{-1}}}
\newcommand{\chidof}{\ensuremath{\chi^2/\mbox{d.o.f.}}}
\newcommand{\dchidof}{\ensuremath{\Delta\chi^2/\Delta\mbox{d.o.f.}}}
\newcommand{\cdof}{\ensuremath{C/\mbox{d.o.f.}}}

\newcommand{\nhsym}{\ensuremath{N_{\mbox{\scriptsize H}}}}
\newcommand{\ha}{\ensuremath{\mbox{H}\alpha}}
\newcommand{\hb}{\ensuremath{\mbox{H}\beta}}
\newcommand{\hea}{\ensuremath{\mbox{He}\alpha}}
\newcommand{\OVII}{\ion{O}{vii}}    
\newcommand{\Lya}{\ensuremath{\mbox{Ly}\alpha}}    
\newcommand{\feka}{\ensuremath{\mbox{Fe~K}\alpha}}
\newcommand{\fekb}{\ensuremath{\mbox{Fe~K}\beta}}
\newcommand{\nika}{\ensuremath{\mbox{Ni~K}\alpha}}
\newcommand{\src}{Mrk~915}
\newcommand{\uno}{{\sc OBS\,1}}
\newcommand{\due}{{\sc OBS\,2}}
\newcommand{\tre}{{\sc OBS\,3}}
\title[\xmm$+$\nustar\ joint observations of \src]{\xmm\ and \nustar\ joint observations of 
\src: 
a deep look into the X-ray properties\thanks{Based on observations obtained with the \xmm\ and the \nustar\ satellites.}}

    \author[L.~Ballo et al.]{
      L.~Ballo,$^{1,2}$\thanks{E-mail: lucia.ballo@gmail.com}
      P.~Severgnini,$^{1}$
      R.~Della~Ceca,$^{1}$
      V.~Braito,$^{3,4}$
      S.~Campana,$^{3}$
      A.~Moretti,$^{1}$
  \newauthor 
      C.~Vignali,$^{5,6}$
      A.~Zaino$^{1,5}$
    \\
   $^{1}$Osservatorio Astronomico di Brera (INAF), via Brera 28, I-20121, Milano (Italy)\\
   $^{2}$XMM-\emph{Newton} Science Operations Centre, ESAC/ESA, PO Box 78, E-28692 Villanueva de la Ca\~{n}ada, Madrid (Spain)\\
   $^{3}$Osservatorio Astronomico di Brera (INAF), via E. Bianchi 46, I-23807 Merate, LC (Italy)\\
   $^{4}$Department of Physics, University of Maryland, Baltimore County, Baltimore, MD 21250 (USA)\\
   $^{5}$Dipartimento di Fisica e Astronomia, Universit\`a degli Studi di Bologna, Via Gobetti 93/2, I-40129, Bologna (Italy)\\
   $^{6}$Osservatorio Astronomico di Bologna (INAF), Via Gobetti 93/3, I-40129, Bologna (Italy)\\
    }

\date{Accepted 2017 May 31. Received 2017 May 31; in original form 2017 February 01}

\pubyear{2017}

\begin{document}
\label{firstpage}
\pagerange{\pageref{firstpage}--\pageref{lastpage}}
\maketitle

\begin{abstract}
We report on the X-ray monitoring programme (covering slightly more than $11\,$days) carried out jointly by \xmm\ 
and \nustar\ 
on the 
intermediate Seyfert galaxy \src.
The light curves extracted in different energy ranges show a variation in intensity but not a significant change in spectral shape.
The X-ray spectra 
reveal the presence of a two-phase warm absorber: 
a fully covering mildly ionized structure [$\log \xi/(\ionpar) \sim 2.3$, 
$\nhsym\sim 1.3\times10^{21}\,$\nh]
and a partial covering ($\sim 90\,$per cent) lower ionized one [$\log \xi/(\ionpar) \sim 0.6$, $\nhsym\sim 2\times10^{22}\,$\nh].
A reflection component from distant matter is also present.
Finally, a high-column density ($\nhsym\sim 1.5\times10^{23}\,$\nh) distribution of neutral matter covering a small fraction 
of the central region is observed, almost constant, in all observations.
Main driver of the variations observed between the datasets is a decrease in the intrinsic emission by a factor of $\sim 1.5$.
Slight variations in the partial covering ionized absorber are detected, while the data are consistent with no variation of
the total covering absorber.
The most likely interpretation of the present data locates this complex absorber closer to the central source than the narrow line region,
possibly in the broad line region, in the innermost part of the torus, or in between.
The neutral obscurer may either be part of this same stratified structure
or associated with the walls of the torus, grazed (and partially intercepting) the line of sight.
\end{abstract}




\section{Introduction}\label{sect:intro}

In the unification paradigm of active galactic nuclei (AGN) by \citet{antonucci93}, types~1 and 2 AGN 
classification is a consequence of
the inclination angle, 
more specifically whether the outer molecular torus, supposed to obscure the accretion disk and the broad line region (BLR), intercepts the line of sight. 
This ``static'' view, in which the torus is made of a smooth dusty distribution, has been questioned in recent years by several studies 
\citep[\eg,][]{elitzur06,nenkova08,honig10,merloni14}.
The emerging picture, due to the
increasing amount of observations at different wavelengths, 
requires 
a more complex structure, maybe clumpy, with multiple absorbers present around the central source at different physical scales 
\citep[and references therein]{bianchi12,netzer15}.

Focusing on the X-ray domain,
measurements of absorption variability are commonly detected in nearby obscured AGN.
These observations imply that the circumnuclear X-ray absorber 
(or, at least, part of it) must be clumpy, and located at sub-parsec distances from the central source \citep[see \eg,][]{bianchi12}, 
requiring the presence of absorbing gas inside the dust sublimation radius.
For NGC~1365, the prototype of AGN showing extreme variations in the X-ray absorbers, the large amount of high-quality data from short-timescales monitoring and 
long observations, allowed to deeply investigate the nature of the absorbing medium.
The derived distances and physical parameters, typical of BLR clouds, 
strongly suggest that the X-ray absorber and the clouds responsible for broad emission lines in the optical/UV are at the same distance from the central black hole 
(see \citealt{risaliti16} for a review of the X-ray observations of NGC~1365).
Recent 
observations of NGC~1365 with \xmm\ 
revealed the presence of a multi-zone warm absorber with different ionization levels
\citep{braito14}.
Indeed, it is nowadays known that
the sub-pc/pc scale region of AGN houses also ``warm'' (i.e., partially photoionized, $T\sim10^4-10^6\,$K) gas that absorbs 
the nuclear emission in the X-ray band. 
Warm absorbers, observed in at least $50$\% of the unobscured AGN and quasars, are often found to be complex and multi-phase
(\eg, \citealt{crenshaw03,piconcelli04,porquet04,blustin05,mckernan07}; see \citealt{costantini10} for a recent review).
Outflowing with velocities
$\lesssim 3000\,$km/s, 
these systems possibly
originate in a thermally driven wind arising from the inner walls of the 
molecular torus \citep{blustin05}, or represent a later stage of an accretion-disk driven wind \citep{proga04}.
The  differently ionized layers of warm absorbers could be part of a single large-scale stratified outflow \citep{tombesi13}. 

The emerging scenario therefore requires dust-free and dusty clumps that coexist in one physical region that contains the BLR, closer in, and the torus, 
further away from the central black hole \citep[a recent review can be found in][]{netzer15}.
Warm gas, possibly outflowing, can contribute to the intercloud BLR medium that provide the pressure confinement 
for the cold BLR clouds \citep[\eg,][]{miniutti14}.
In order to investigate the interconnection between these different structures, a powerful tool is to perform X-ray monitoring of AGN observed at 
intermediate angles, 
where we expect that the line of sight grazes the torus.

\src\ is a nearby galaxy ($z=0.024$) optically classified as Seyfert~1.5 - Seyfert~1.9, depending on the intensity/presence of the broad components 
of the \ha\ and
\hb\ observed at different epochs in the optical spectra \citep{goodrich95,bennert06,trippe10}.
As summarized in \citet{severgnini15}, this spectral variability could be due to a change in the reddening level, to a change
in the nuclear photoionizing continuum, or to a combination of both.
Independently of the origin of this spectral variability, at least part of the optical absorption affecting the central regions originates outside the obscuring
torus, and it is most likely associated with the dust lanes seen crossing the central source \citep{malkan98,munoz07}.
The stellar disc inclination derived by near-ultraviolet imaging \citep[$35\arcdeg-57\arcdeg$;][]{munoz07} implies that in \src\ our line of sight 
is only partially blocked by the torus,
assumed here to be coplanar with the galaxy plane, and with an half-opening angle of $\sim 60\arcdeg$.
Moreover, in \src\ the torus is likely to have a clumpy structure, as suggested by observations with the Infrared Spectrograph (IRS) on board the 
\spitzer\ Space Telescope \citep{mendoza15}.

In the X-ray, archival \swift\ X-ray Telescope \citep[XRT;][]{xrt} data strongly suggested the presence of a variable X-ray obscuring medium \citep{ballo14}. 
A daily XRT monitoring programme spanning $\sim 3\,$weeks, presented in \citet{severgnini15}, allowed us to confirm this hypothesis.
We detected a significant count rate 
variation on a time-scale of a few days, due to a change by a factor $\sim 1.5$ in the intrinsic nuclear power coupled with
a change in the ionization state and/or in the covering factor of a partial covering ionized absorber.
The analysis of a $\sim 35\,$ksec \suzaku\ observation of \src\ supports the presence of ionized obscuration \citep{kawamuro16}.

In this paper, we present the results from a simultaneous \nustar\ and \xmm\ monitoring campaign (covering the $0.3-70\,$keV energy band) 
performed in 2014 December.
The paper is organized as follows. 
In Section~\ref{sect:obs} we discuss the X-ray observations and data reduction, while a first comparison between the datasets is presented in 
Section~\ref{sect:lc}. 
The X-ray spectral analysis is described in Section~\ref{sect:xray}, with Sections~\ref{sect:rgs} and \ref{sect:bb} devoted to the high energy resolution data 
and 
the broadband spectra, respectively.
In Section~\ref{sect:disc} we discuss our results, comparing them with the previous X-ray results, and finally in Section~\ref{sect:concl} 
we summarize our work. 
Throughout the paper we assume a flat $\Lambda$CDM cosmology with $\pedix{H}{0}=71\,$\kms~Mpc$^{-1}$, 
$\Omega_\Lambda=0.7$ and $\pedix{\Omega}{M}=0.3$.

%
\begin{table*}
\begin{minipage}[t]{1\textwidth}
 \caption{\xmm\ and \nustar\ observation log for \src.}
 \label{tab:log}
 \begin{center}
{
\footnotesize
  \begin{tabular}{@{\extracolsep{0cm}}r@{\extracolsep{0.5cm}} l@{\extracolsep{0.2cm}} l@{\extracolsep{0.2cm}} r@{\extracolsep{0.cm}-}c@{\extracolsep{0.cm}-}l@{\extracolsep{0.1cm}} r@{\extracolsep{0.cm}:}c@{\extracolsep{0.cm}:}l@{\extracolsep{0.2cm}} r@{\extracolsep{0.cm}-}c@{\extracolsep{0.cm}-}l@{\extracolsep{0.1cm}} r@{\extracolsep{0.cm}:}c@{\extracolsep{0.cm}:}l@{\extracolsep{0.2cm}} c@{\extracolsep{0.2cm}} c@{\extracolsep{0.2cm}} c@{}}
  \hline  \hline
Obs. & Satellite & Obs. ID & \multicolumn{6}{c}{Start} & \multicolumn{6}{c}{Stop} & Detector & Net count rate$^{a}$  &  Net exp. time$^{b}$\\
   &   &   & \multicolumn{6}{c}{} & \multicolumn{6}{c}{} &  &  [counts~s$^{-1}$]  &  [ks]
 \vspace{0.1cm} \\
 \hline
 \vspace{-0.2cm} \\
   & \nustar\ & 60002060002 & 2014 & 12 & 02 & 13 & 56 & 07 & 2014 & 12 & 03 & 18 & 46 & 07 & FPMA & $0.216 \pm 0.002$ & $53.0$ \\
   &   &   & \multicolumn{6}{c}{}   &  \multicolumn{6}{c}{}  & FPMB & $0.213 \pm 0.002$ & $52.9$ \\
   & \xmm\ & 0744490401 & 2014 & 12 & 02 & 13 & 54 & 36 & 2014 & 12 & 04 & 02 & 29 & 18 & pn & $1.183 \pm 0.004$ & $88.9$ \\
   \uno\ &  &  & 2014 & 12 & 02 & 14 & 09 & 24 & 2014 & 12 & 04 & 01 & 48 & 22 & MOS1 & $0.400 \pm 0.002$ & $100.2$ \\
   &   &  & 2014 & 12 & 02 & 13 & 26 & 16 & 2014 & 12 & 04 & 02 & 33 & 29 & MOS2 & $0.398 \pm 0.002$ & $102.0$ \\
   &   &  & 2014 & 12 & 02 & 13 & 25 & 38 & 2014 & 12 & 03 & 22 & 03 & 56 & RGS1-S$^{c}$ & $0.0121 \pm 0.0006$ & $87.5$ \\
   &   &  & 2014 & 12 & 03 & 22 & 31 & 52 & 2014 & 12 & 04 & 02 & 37 & 33 & RGS1-U$^{c}$ & $0.008 \pm 0.002$ & $12.3$ \\
   &   &  & 2014 & 12 & 02 & 13 & 25 & 46 & 2014 & 12 & 04 & 02 & 37 & 36 & RGS2 & $0.0134 \pm 0.0005$ & $101.1$ 
 \vspace{0.1cm} \\
 \hline
 \vspace{-0.2cm} \\
   & \nustar\ & 60002060004 & 2014 & 12 & 07 & 06 & 51 & 07 & 2014 & 12 & 08 & 12 & 46 & 07 & FPMA & $0.149 \pm 0.002$ & $54.3$ \\
   &   &   & \multicolumn{6}{c}{}   &  \multicolumn{6}{c}{}  & FPMB & $0.145 \pm 0.002$ & $54.1$ \\
   & \xmm\ & 0744490501 & 2014 & 12 & 07 & 08 & 32 & 29 & 2014 & 12 & 08 & 02 & 52 & 10 & pn & $0.807 \pm 0.004$ & $41.4$ \\
   \due\ &  &  & 2014 & 12 & 07 & 08 & 13 & 41 & 2014 & 12 & 08 & 02 & 56 & 17 & MOS1 & $0.269 \pm 0.002$ & $51.1$ \\
   &   &  & 2014 & 12 & 07 & 08 & 04 & 10 & 2014 & 12 & 08 & 02 & 56 & 22 & MOS2 & $0.276 \pm 0.002$ & $49.2$ \\
   &   &  & 2014 & 12 & 07 & 08 & 03 & 31 & 2014 & 12 & 08 & 03 & 00 & 28 & RGS1 & $0.0084 \pm 0.0008$ & $49.7$ \\
   &   &  & 2014 & 12 & 07 & 08 & 03 & 39 & 2014 & 12 & 08 & 03 & 00 & 22 & RGS2 & $0.0090 \pm 0.0007$ & $49.7$ 
 \vspace{0.1cm} \\
 \hline
 \vspace{-0.2cm} \\
   & \nustar\ & 60002060006 & 2014 & 12 & 12 & 12 & 41 & 07 & 2014 & 12 & 13 & 17 & 01 & 07 & FPMA & $0.118 \pm 0.002$ & $50.7$ \\
   &   &   & \multicolumn{6}{c}{}   &  \multicolumn{6}{c}{}  & FPMB & $0.106 \pm 0.002$ & $50.6$ \\
   & \xmm\ & 0744490601 & 2014 & 12 & 12 & 13 & 12 & 30 & 2014 & 12 & 13 & 04 & 17 & 13 & pn & $0.718 \pm 0.006$ & $22.6$ \\
   \tre\ &  &  & 2014 & 12 & 12 & 12 & 43 & 41 & 2014 & 12 & 13 & 04 & 21 & 19 & MOS1 & $0.240 \pm 0.003$ & $31.9$ \\
   &   &  & 2014 & 12 & 12 & 12 & 44 & 12 & 2014 & 12 & 13 & 04 & 21 & 24 & MOS2 & $0.243 \pm 0.003$ & $32.8$ \\
   &   &  & 2014 & 12 & 12 & 12 & 43 & 32 & 2014 & 12 & 13 & 04 & 25 & 25 & RGS1 & $0.007 \pm 0.001$ & $27.8$ \\
   &   &  & 2014 & 12 & 12 & 12 & 43 & 40 & 2014 & 12 & 13 & 04 & 25 & 26 & RGS2 & $0.008 \pm 0.001$ & $27.6$ \\
  \end{tabular}
 }
  \end{center}       
 {\footnotesize   {\sc Note:} $^{a}\,$Net source count rate after screening and background subtraction, as observed in the $0.3-10\,$keV (pn, MOS1 and MOS2), 
 $0.5-2\,$keV (RGS1 and RGS2), and $4-70\,$keV (FPMA and FPMB) energy ranges.
 $^{b}\,$ Net exposure time, after screening was applied on the data.
 $^{c}\,$ S$=$scheduled; U$=$unscheduled.

 }
\end{minipage}
\end{table*}
%

\section{Observation and data reduction}\label{sect:obs}

\subsection{\xmm}\label{sect:xmm}

We observed \src\ with \xmm\ in 2014 December in three pointings (hereafter, \uno, \due, and \tre; see Table~\ref{tab:log}), 
separated from each other by about $5$ days, for a total of about $230\:\,$ksec 
(Obs.~ID $0744490401$, $0744490501$ and $0744490601$). 

The observations were performed with the European Photon Imaging Camera (EPIC), the Optical Monitor (OM) and the 
Reflection Grating Spectrometer (RGS).
In this paper, we concentrate on the data in the X-ray band (EPIC and RGS spectra).
The three EPIC cameras \citep[pn, MOS1, and MOS2;][]{pn,mos} were operating in large window mode, with the thin filter applied.

The \xmm\ data have been processed and cleaned using the Science Analysis Software ({\sc SAS} version~14.0) with the most recent calibrations; 
the tasks {\sc epproc} and {\sc emproc} were run to produce calibrated and concatenated event lists for the EPIC cameras.
EPIC event files have been filtered for high-background time intervals, following the standard method consisting in rejecting
time periods of high count rate at energies $>10\,$keV. 
A strong flare of $\sim21\,$ ksec is observed during \uno, starting at $\sim 38\,$ksec after the beginning of the observation.

Events corresponding to patterns $0-12$ (MOS1\&2) and $0-4$ (pn) have been used.
From the observed count rate, we can exclude event pileup in the EPIC data (see Table~\ref{tab:log}).
The spectral response matrices at the source position were generated using the {\sc SAS} tasks {\sc arfgen} and {\sc rmfgen}
and the latest calibration files.

The EPIC source counts were extracted from a circular region of radius $30\arcsec$ (corresponding to an encircled energy fraction of
$\sim90\,$per cent); background counts were extracted from two source-free circular regions 
in the same chip of $30\arcsec$ radius each.
Spectra were re-binned with the task {\sc specgroup}, to not over-sample the instrument energy resolution by more than a factor of three, 
and to have at least $50$ counts in each energy bin.
This allows the application of $\chi^2$ statistics.

The RGS \citep{rgs} data have been reduced using the standard {\sc SAS} task {\sc rgsproc}, and the most recent calibration files; 
details on net exposure times and total counts 
are reported in Table~\ref{tab:log}.
After filtering out the high-background time intervals, 
for each observation we combined the RGS1 and RGS2 spectra (found to be  in good agreement, typically to within the $3$\% level)
using the  {\sc SAS} task {\sc rgscombine};
the spectra were binned at $\Delta\lambda=0.1\,$\AA\ (slightly undersampling the FWHM spectral resolution).

\subsection{\nustar}\label{sect:nus}

\nustar\ \citep{nustar} observed \src\ simultaneously with \xmm\ with its two co-aligned telescopes with corresponding Focal Plane Modules A 
(FPMA) and B (FPMB) for a total of about $310\:\,$ksec  of elapsed time. 
The level 1 data products were processed with the \nustar\ Data Analysis Software ({\sc NuSTARDAS}) package (ver. 1.4.1). 
Event files (level 2 data products) were produced, calibrated, and cleaned using standard filtering criteria with the {\sc nupipeline} task 
and the latest calibration files available in the NuSTAR calibration database ({\sc CALDB}). 
The FPMA and FPMB source counts were extracted from a circular region of radius $60\arcsec$; this aperture approximates the $\approx 90$\% 
encircled-energy fraction contour of the point-spread function.
Background counts were extracted from two source-free circular regions of $60\arcsec$ radius each.
\nustar\ detects the source above $30\,$keV with a signal-to-noise ratio of $\sim 13$, $\sim 9$ and $\sim 7$ 
in \uno, \due, and \tre, respectively.
Spectra were binned in order to have at least $50$ counts in each energy bin.
Exposure times and total count rates for each spectrum 
are reported 
in Table~\ref{tab:log}.

\section{X-ray variability: a first comparison between the datasets}\label{sect:lc}

   \begin{figure}
   \centering
    \resizebox{1.\hsize}{!}{\includegraphics{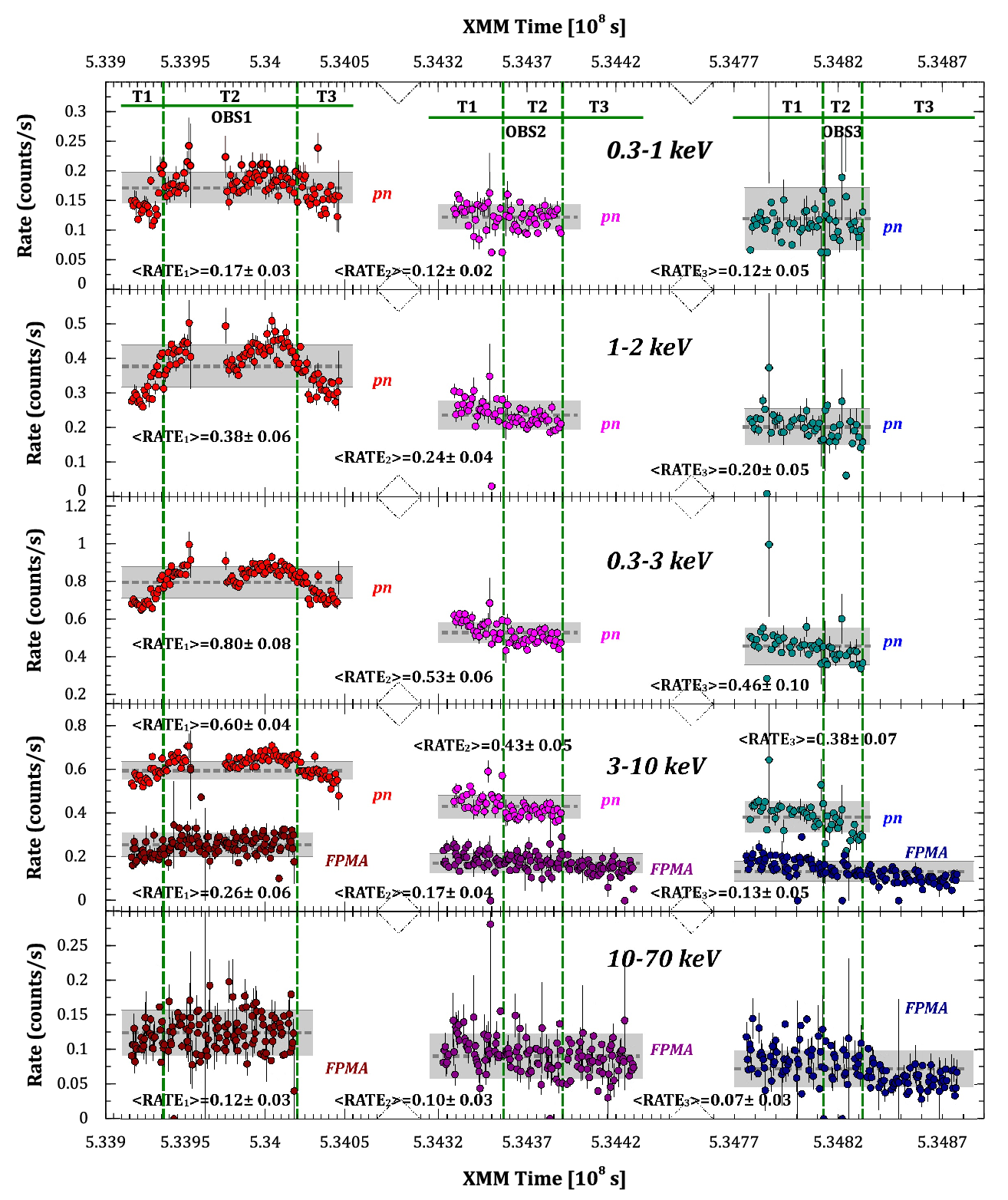}}
    \resizebox{1.\hsize}{!}{\includegraphics{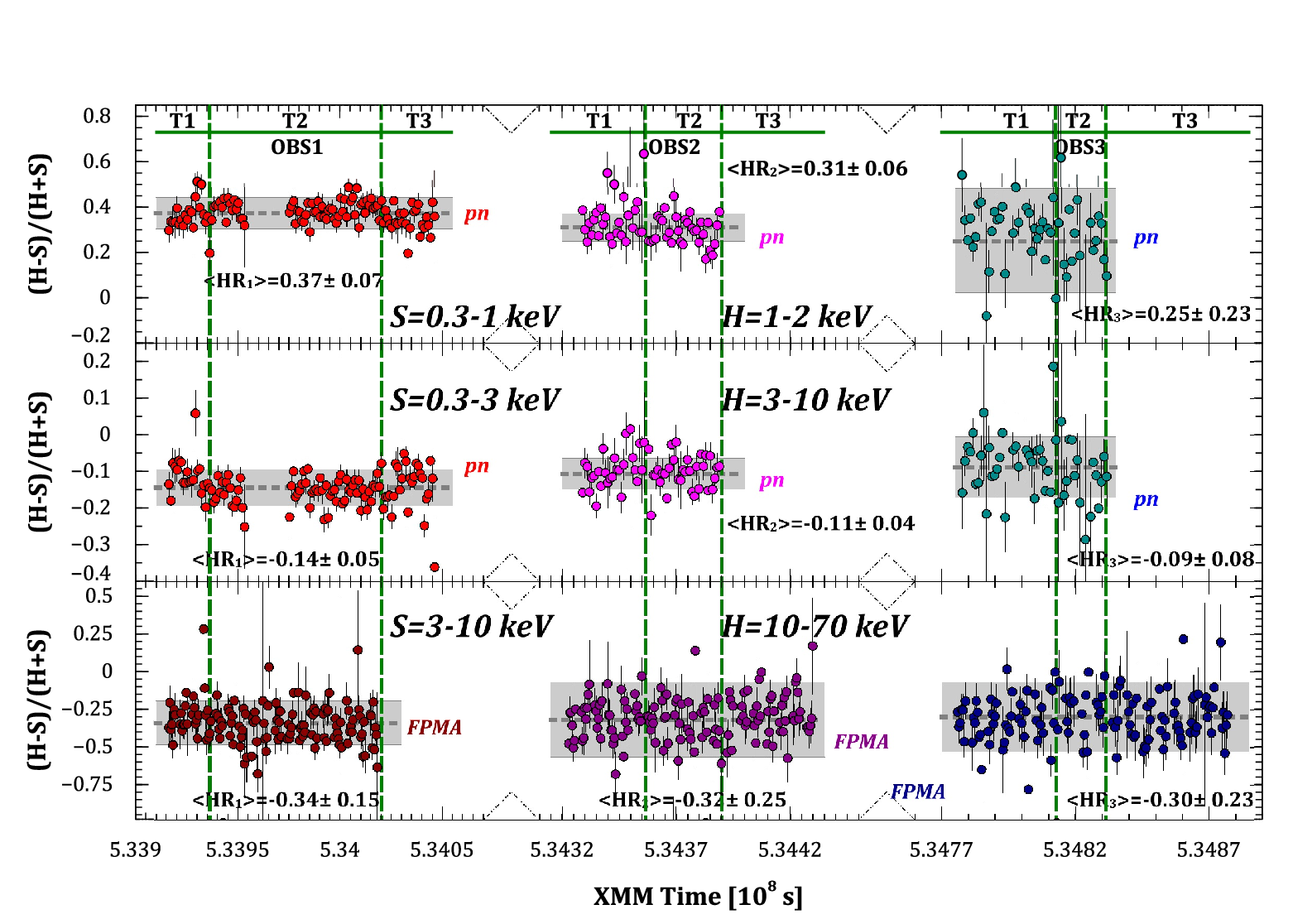}}
   \caption{Background-subtracted light curves of \src\ from the \xmm\ pn and the \nustar\ FPMA observations (labeled as ``\uno'', ``\due'' and ``\tre''; 
   green horizontal lines). 
   Similar results are obtained if we consider the \nustar\ camera FPMB instead of FPMA, and/or the EPIC-MOS data instead of the EPIC-pn.
   For each dataset, we report the mean rate (grey horizontal dashed line) and the standard deviation (shaded area) for the whole observation.
   The vertical green dashed lines mark the intervals (T1, T2 and T3) in which each observation has been divided (see Sect.~\ref{sect:lc} and 
   Sect.~\ref{sect:xray}).
   {\it Up:} Count rates. From top to bottom: $0.3-1\,$keV, $1-2\,$keV and  $0.3-3\,$keV (pn only); $3-10\,$keV (both cameras); $10-70\,$keV (FPMA only).
   {\it Down:} Hardness ratios HR=[RATE(H)-RATE(S)]/[RATE(H)+RATE(S)]. 
   Upper panel: $\mbox{H}=1-2\,$keV and $\mbox{S}=0.3-1\,$keV (pn only); middel panel: $\mbox{H}=3-10\,$keV and $\mbox{S}=0.3-3\,$keV (pn only); 
   lower panel: $\mbox{H}=10-70\,$keV and $\mbox{S}=3-10\,$keV (FPMA only).
      }
              \label{fig:lc}%
    \end{figure}

We tested time-variability within the \xmm\ and \nustar\ observations, generating
source light curves in several energy intervals with
a binning time of $1000\,$s (EPIC cameras) and $500\,$s (FPMs).
Fig.~\ref{fig:lc} (upper panels) shows the X-ray count rate light curves obtained with the EPIC-pn and FPMA 
in the $0.3-1\,$keV, $1-2\,$keV, $0.3-3\,$keV, $3-10\,$keV and $10-70\,$keV energy ranges.
In the lower panels, we present the 
hardness ratios (HRs) defined as HR=[RATE(H)-RATE(S)]/[RATE(H)+RATE(S)], 
where RATE(H) and RATE(S) are the count rates observed in the hard and soft bands, respectively.
Similar results are obtained if we consider the \nustar\ camera FPMB instead of FPMA, and/or the EPIC-MOS data instead of the EPIC-pn.
Variations at $>99.9\,$per cent confidence level ($\chi^2$ test)
are observed during \uno\ and \due\ in the \xmm\ 
data in all intervals above $\sim 1\,$keV; 
variations observed during \tre\ are significant at less than $99.9\,$per cent confidence level.
The \nustar\ data do not show statistically significant variations, except for observations taken during the \tre\ period.

In the following analysis, each observation has been divided in three time intervals
(labeled as T1, T2 and T3 in Fig.~\ref{fig:lc}), as detailed below.
\uno\ shows the most complex pattern, with a $\sim 35\,$per cent [$\sim 25\,$per cent] rise in flux observed by \xmm\ 
between $0.3$ and $10\,$keV [by \nustar\ between $3$ and $70\,$keV] 
during the first $\sim 25\,$ksec (time interval marked as \uno-T1 in Fig.~\ref{fig:lc}).
Low-level variations (less than $10\,$per cent) are observed at time from $\sim 25$ to $\sim 70\,$ksec (\uno-T2 interval) both by \xmm\ and \nustar.
After that, the flux lowers until the end of the \xmm\ observation (\uno-T3 interval), recovering the initial intensity.

Both \due\ and \tre\ are characterized by a smooth decline over the whole observation, with a $\sim 21\,$per cent and $\sim 28\,$per cent drop in the
\xmm\ flux, respectively.
In these two observations, the separation between T1 and T2 was selected according to a straight line fitted to the count rate observed by
\xmm\ in the $0.3-10\,$keV energy range being higher or lower than the mean count rate.
The dividing time between T2 and T3 corresponds to the end of the \xmm\ observation 
(see Fig.~\ref{fig:lc}).
\nustar, observing for longer time intervals, shows a decreasing of $\sim 37\,$per cent and $\sim 41\,$per cent in the flux.

However, as shown in the lower panels of Fig.~\ref{fig:lc}, the HR light curves show a lack of significant spectral variability over each observation 
in each energy band. 

Variations between the three observations are clearly visible in all bands, with a drop of $\sim 30\,$per cent between the mean count rates observed in 
\uno\ and \due.
A low-level decrease in the mean count rate is observed comparing \due\ and \tre, with a change of $\sim 13\,$per cent [$\sim 20\,$per cent] 
in \xmm\ [\nustar].
Again, the mean HRs are consistent within the errors (see Fig.~\ref{fig:lc}).

\section{X-ray spectral analysis}\label{sect:xray}

In this Section we present the analysis of the X-ray spectra of \src.
We first discuss the high-energy resolution data from the RGS, and then we move to the analysis of the \xmm-EPIC$+$\nustar\ spectra.
Our study of the light curves suggests that during each observation the source vary mainly in flux, since the HR light curves show a lack of significant 
spectral variability; therefore,
to increase the statistics, in the spectral analysis for each observation we consider the time-averaged data over the whole duration.
As for the broad-band spectral analysis only, the results obtained on each observation are then compared with the spectra extracted in the relevant 
T1, T2 and T3 intervals (see Section~\ref{sect:lc}).

Spectral fits were performed 
using the X-ray spectral fitting package XSPEC \citep{xspec} v12.9.0.
All the models discussed in the following assume Galactic absorption with a column density of $\pedix{N}{H,Gal}=5.3\times 10^{20}\,$\nh\ \citep{nh}.
To model both Galactic and intrinsic absorptions we used the {\sc (z)tbabs} model in XSPEC, adopting cross-sections and abundances of \citet{wilms}.

   \begin{figure*}
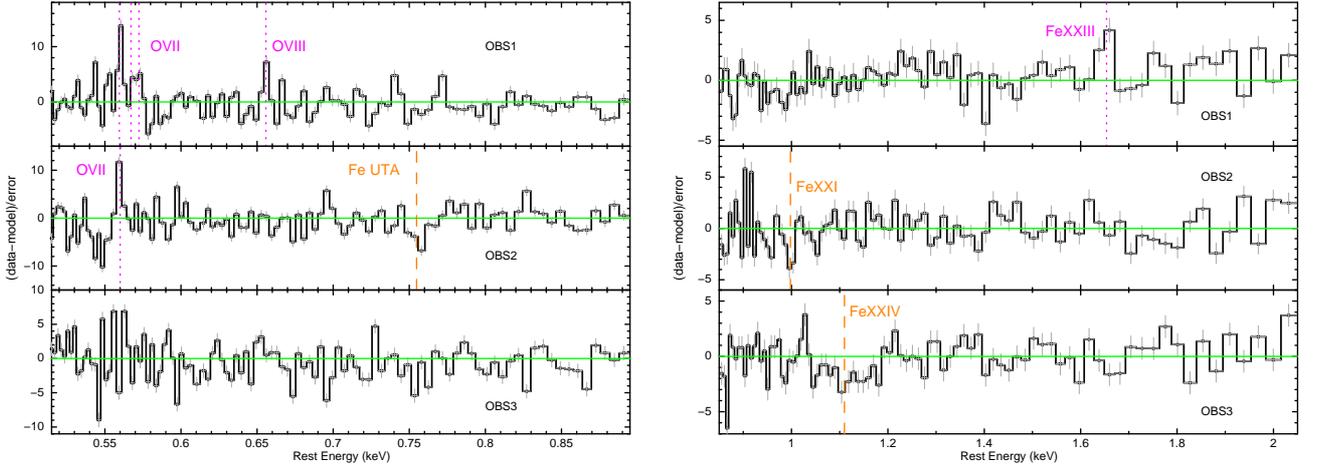

   \centering
    \resizebox{0.49\hsize}{!}{\includegraphics[angle=270]{f2a.ps}}
    \resizebox{0.49\hsize}{!}{\includegraphics[angle=270]{f2b.ps}}
   \caption{Residuals in the RGS \uno\ (top), \due\ (middle), and \tre\ (bottom) spectra against the continuum model 
   (from \citealt{severgnini15}, applied to the EPIC$+$\nustar\ data).
   The data are plotted in the source rest frame, between $0.5$ and $0.9\,$keV (left), and between $0.85$ and $2\,$keV (right).
   The vertical lines mark the positions of the emission (dotted magenta lines) and absorption (orange dashed lines) features detected 
   during the analysis.
    }
              \label{fig:rgsde}%
    \end{figure*}
%

   \begin{figure*}
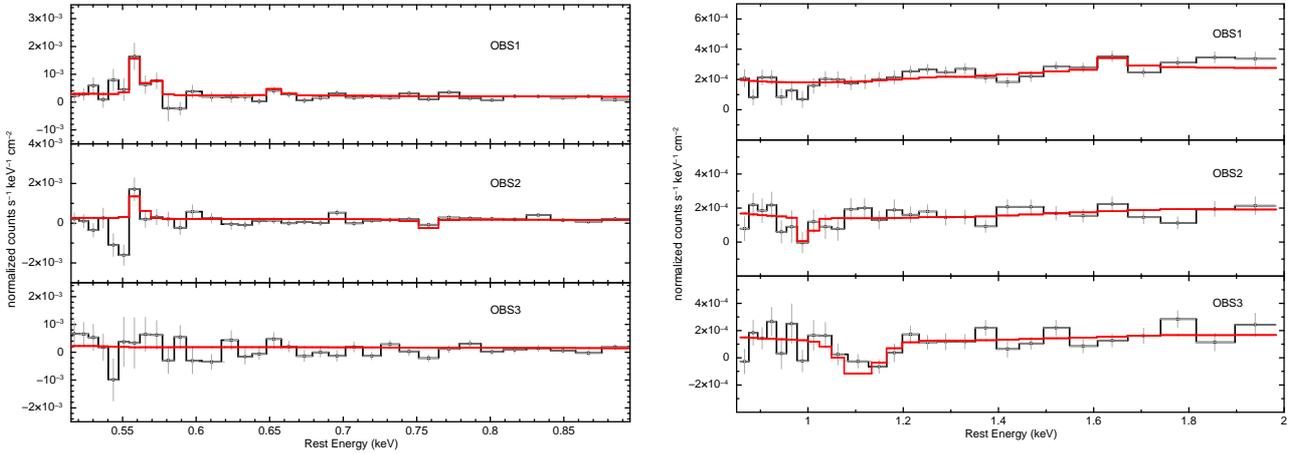

   \centering
    \resizebox{0.49\hsize}{!}{\includegraphics[angle=270]{f3a.ps}}
    \resizebox{0.49\hsize}{!}{\includegraphics[angle=270]{f3b.ps}}
   \caption{RGS spectra (\uno, top; \due, middle; \tre, bottom) fitted with the model including an absorbed power law continuum with the detected 
   absorption and emission lines 
   superimposed.
   The data are plotted in the source rest frame, between $0.5$ and $0.9\,$keV (left), and between $0.85$ and $2\,$keV (right).
   }
              \label{fig:rgslda}%
    \end{figure*}
%

\subsection{High-resolution RGS data}\label{sect:rgs}

As a first step, we concentrated on the spectral analysis of the RGS data.
The spectra were 
analyzed over the $0.5-2.0\,$keV energy range, with the $C$-statistic applied \citep{cstat}.

The narrow RGS energy range
prevents us from performing a detailed spectral fitting of the continuum;
an initial guess for its shape was obtained by applying the model describing the \swift-XRT data \citep{severgnini15}
to the EPIC ($0.3-10\,$keV) $+$\nustar\ ($4-70\,$keV) data\footnote{The adopted model is a partially absorbed power law 
($\nhsym\sim2.5\times10^{22}\,$\nh, covering factor $\sim85\,$per cent).}.
In Fig.~\ref{fig:rgsde} we show the residuals in the RGS data against the resulting continuum.
Line-like structures both in emission (at $\sim 0.55-0.6\,$keV and $\sim1.6\,$keV) and in absorption 
(at $\sim0.75\,$keV and $\sim0.9-1.2\,$keV), variable between the observations, are visible.

In order 
to test the 
significance
of individual features, for each observation 
we divided the energy ranges on narrow intervals of $\Delta E\sim0.2-0.3\,$keV, 
and subsequently added narrow Gaussian lines to the model; 
the positions and the normalizations were left free to vary,
while the line widths were initially fixed to $2\,$eV and subsequently (where necessary) were allowed to change.
At each step, we also allowed the normalization 
of the continuum model to vary, while we fixed the parameters of the previously detected lines.
The detection of an individual line was considered statistically significant if its addition to the previous model (composed of the 
absorbed power law
with the addition of all the previously detected lines) resulted in an improvement of the fit statistic of 
$\Delta C>9.2$ (corresponding to $99$\% significance level
for two parameters of interest), when evaluated in the energy interval considered for the fit. 
We finally fitted the model composed by the baseline continuum (the partially absorbed power law) plus all the detected absorption and emission lines, 
over the whole energy range, 
allowing all the line energies and normalizations to vary, as well as the continuum shape and intensity.

Table~\ref{tab:bfrgs} lists the rest-frame energy of the emission (first part) and absorption lines (second part) 
detected, along with their intensity and the likely identification of the atomic transition
(corresponding to the label that identifies each line in Fig.~\ref{fig:rgsde}).
The last three columns list 
the statistical improvement,
the order of inclusion in the model of each line,
and the energy interval considered for evaluating the significance of the feature.
The final model is shown in Fig.~\ref{fig:rgslda}.

The observed absorption lines do not show any significant shift in the energy centroid with respect to the known atomic energy.
An upper limit on the outflow velocity of $\pedix{v}{gas}\lesssim 3000\,\kms$ can be derived from the $90\,$per cent errors on the energy centroids.
Multiple absorption components may be required in order to model the wide range of ionization states
of the gas, covering the Fe Unresolved Transmission Array (UTA) as well as the \ion{Fe}{xxi} and higher.
An absorption trough likely associated to \ion{Fe}{xvii-xviii} was tentatively detected also in the \swift-XRT spectra by \citet{severgnini15}.

%
\begin{table*}
\begin{minipage}[t]{1\textwidth}
\caption{Combined \xmm-RGS1 and RGS2: best-fit absorption and emission lines required.}
\label{tab:bfrgs}
\begin{center}
{
\footnotesize
\begin{tabular}{@{\extracolsep{0cm}}c@{\extracolsep{0.4cm}} c@{\extracolsep{0.13cm}} c@{\extracolsep{0.18cm}} c@{\extracolsep{0.18cm}} c@{\extracolsep{0.18cm}} c@{\extracolsep{0.18cm}} c@{\extracolsep{0.13cm}} c@{\extracolsep{0.13cm}} c}
 \hline \hline
 \multicolumn{9}{c}{\sc Emission Lines} \\
\vspace{-0.2cm}\\
    \hline         
\vspace{-0.2cm}\\
    Obs.
    & Rest E
    & Intensity
    & $\sigma$
    & ID
    & Atomic E
    & $\Delta C$
    & Sequence
    & $\Delta \pedix{E}{fit}$ \\
   (1)
   & (2)
   & (3)
   & (4)
   & (5)
   & (6) 
   & (7)
   & (8)
   & (9) \\
\vspace{-0.2cm}\\
    \hline         
\vspace{-0.2cm}\\
    \uno\ & $0.560\pm 0.001$ & $2.95\errUD{1.71}{1.59}$ & $2^f$ & \ion{O}{vii}\,\hea\,(f) & $0.561$ & $18.1^{\mydag}$ & I & $0.5-0.7$ \\
     & $0.568^t$ & $<1.33$ & $2^f$ & \ion{O}{vii}\,\hea\,(i) & $0.569$ &  &  & \\
     & $0.573^t$ & $<1.97$ & $2^f$ & \ion{O}{vii}\,\hea\,(r) & $0.574$ &  &  & \\
     & $0.656\pm 0.002$ & $0.56\errUD{0.38}{0.36}$ & $2^f$ & \ion{O}{viii}\,\Lya & $0.653$ & $9.8$ & II & $0.6-0.8$ \\ 
     & $1.654\errUD{0.058}{0.017}$ & $0.64\errUD{0.51}{0.54}$ & $2^f$ & \ion{Fe}{xxiii} & 1.659 & $9.4$ & III & $1.5-1.8$ \\ 
\vspace{-0.15cm}\\
    \due\ & $0.560\pm 0.002$ & $2.85\errUD{2.04}{1.76}$ & $2^f$ & \ion{O}{vii}\,\hea & $0.561-0.574$ & $11.7$ & I & $0.5-0.7$ \\
\vspace{0.2cm}\\
    \hline \hline        
\vspace{-0.2cm}\\
   \multicolumn{9}{c}{\sc Absorption Lines} \\
\vspace{-0.2cm}\\
    \hline         
\vspace{-0.2cm}\\
    Obs.
    & Rest E
    & Intensity
    & $\sigma$
    & ID
    & Atomic E
    & $\Delta C$
    & Sequence
    & $\Delta \pedix{E}{fit}$ \\
   (1)
   & (2)
   & (3)
   & (4)
   & (5)
   & (6) 
   & (7)
   & (8)
   & (9) \\
\vspace{-0.2cm}\\
    \hline         
\vspace{-0.2cm}\\
    \due\ & $0.755\errUD{0.007}{0.003}$ & $-\left(1.14\errUD{1.64}{0.89}\right)$ & $2^f$ &  Fe UTA & $0.729-0.775$ & $9.4$ & III & $0.65-0.85$ \\
     & $0.998\errUD{0.010}{0.009}$ & $-\left(0.69\errUD{1.82}{0.40}\right)$ & $<32$ & \ion{Fe}{xxi} & $0.995$ & $9.3\myddag$ & II & $0.95-1.1$ \\ 
\vspace{-0.15cm}\\
    \tre\ & $1.110\errUD{0.026}{0.015}$ & $-\left(3.06\errUD{3.86}{1.47}\right)$ & $<34$ & \ion{Fe}{xxiv} & $1.110$ & $11.2\myddag$ & I & $1-1.3$ \\ 
  \end{tabular}
}
 \end{center}       
 {\footnotesize   {\sc Note:} 
 The statistic for the best-fit model [for the model with no lines] is $\cdof = 187.4/174$ [$\cdof = 217.2/182$], 
 $\cdof = 206.4/173$ [$\cdof = 284.1/180$] and $\cdof = 192.4/179$ [$\cdof = 203.1/181$], for \uno, \due\ and \tre, respectively. 
 $f=\:$parameter fixed during the fit. 
 $t=\:$tied to the energy of the \OVII\,\hea\,forbidden component ($\pedix{E}{r}-\pedix{E}{f}=0.013\,$keV and $\pedix{E}{i}-\pedix{E}{f}=0.008\,$keV).
 $\mydag\;\Delta\mbox{d.o.f}=4$.
 $\myddag\;\Delta\mbox{d.o.f}=3$.
 \footnotesize Column (1): observation.
 \footnotesize Column (2): measured line energy in the source rest frame, in units of keV.
 \footnotesize Column (3): line normalization, in units of $10^{-5}\,$\normGAUSS.
 \footnotesize Column (4): line width, in units of eV.
 \footnotesize Column (5): possible identification; forbidden, intercombination, and resonance lines denoted by (f), (i), and (r).
 \footnotesize Column (6): known atomic energy of the most likely identification of the line, in units of keV.  
 \footnotesize Column (7): improvement in $C$-statistic (considering only the bins in the energy interval of $\Delta E\sim0.2-0.3\;$keV 
 around the line centroid quoted
 in Column~8) upon adding the line ($\Delta\mbox{d.o.f}=2$) with respect to the baseline model with the addition of the lines detected in the previous steps.
 \footnotesize Column (8): detection sequence.
 \footnotesize Column (9): Energy range considered for the fit, in units of keV.
  }
\end{minipage}
\end{table*}
%

\subsection{Broad-band spectral analysis}\label{sect:bb}

   \begin{figure*}
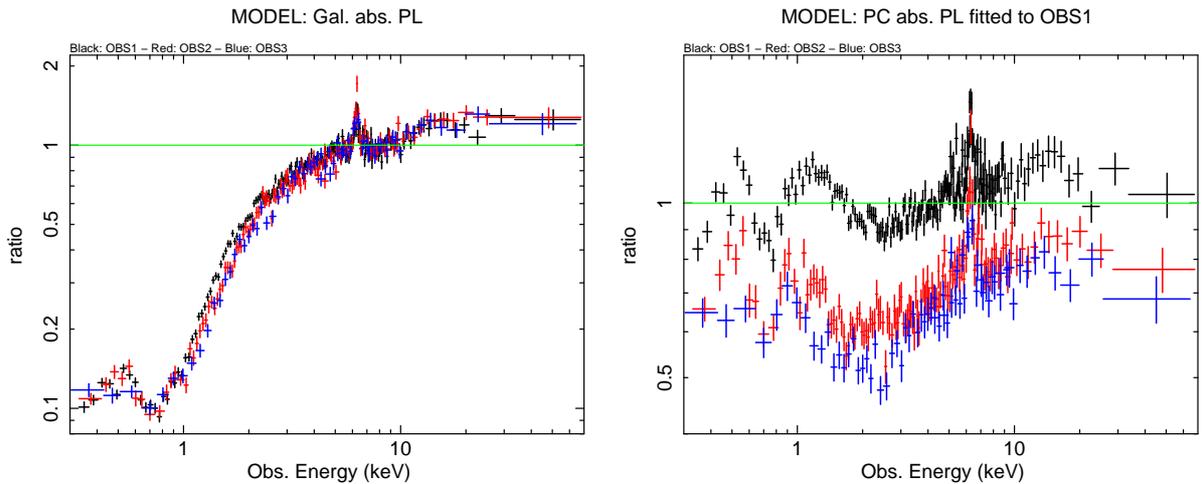

   \centering
      \resizebox{0.45\hsize}{!}{\includegraphics[angle=270]{f4a.ps}}
      \resizebox{0.45\hsize}{!}{\includegraphics[angle=270]{f4b.ps}}
   \caption{Data-to-model ratio for the \xmm\ pn and \nustar\ FPMA spectra: black, \uno; red, \due; blue, \tre.
    All data are rebinned for clarity.
   {\it Left:} the model is a power law, fitted to the data between $5$ and $10\,$keV ($\Gamma\sim 1.53$), 
   absorbed by our Galaxy.
   {\it Right:} a partial covering neutral absorber is added to the model and fitted to the \uno\ data considering the whole energy range. 
   }
              \label{fig:ra}%
    \end{figure*}

As anticipated above, 
in the broad-band spectral analysis we considered the time-averaged data over the whole duration; the results of the analysis 
are then compared with the spectra extracted in the T1, T2, and T3 intervals, as defined in Sect.~\ref{sect:lc}.

In the broad-band spectral analysis, we kept the FPMA and FPMB spectra separate (following the 
prescription\footnote{http://www.nustar.caltech.edu/page/response\_files; see also 
https://heasarc.gsfc.nasa.gov/docs/nustar/analysis, ``\nustar\ Analysis Caveats''}), 
while the MOS1 and MOS2 data have been combined together.
The MOS, pn, FPMA and FPMB spectra have been fitted
simultaneously, keeping the relative normalizations free.
The cross-normalization values between EPIC and FPMs were never larger than a few percent, typically of $3-5$\%, as expected \citep{madsen15b}.
We performed the spectral analysis of the EPIC and \nustar\ spectra over the energy range from $0.3\,$keV to $10\,$keV, 
and from $4\,$keV to $70\,$keV, respectively.
Uncertainties are quoted at the $90$\% confidence level for one parameter of interest ($\Delta\chi^2 = 2.71$). 

In Fig.~\ref{fig:ra} (left panel) we present the \uno\ (black), \due\ (red) and \tre\ (blue) EPIC-pn and FPMA data, shown as a ratio
to a power law ($\Gamma\simeq 1.53$) fitted between $5\,$keV and $10\,$keV.
The power law  normalizations between the observations have been left free to vary.
Considering that the MOS and FPMB spectra agree within the errors with the pn and FPMA spectra, respectively, to avoid clutter only
data from the last cameras are shown.
The model is absorbed by our Galaxy.
This plot clearly shows that there is no significant change in the spectral shape between the three observations.

We started the analysis of the broad-band spectra of \src\ with the longest observation of our monitoring, \uno.
The \due\ and \tre\ datasets were then analysed taking into account the results of the analysis done for the \uno\ spectra.

The strong decrease in the flux below $\sim5\,$keV (see Fig.~\ref{fig:ra}, left panel), particularly enhanced between $0.6$ and $2\,$keV, 
implies the presence of a significant absorption 
intrinsic to the source.
The addition of a neutral absorbing component,
both fully or partially covering the nuclear emission, is not able to reproduce the observed spectral shape: 
in Fig.~\ref{fig:ra} (right panel) we show the data-to-model ratio obtained by fitting over the whole energy range the \uno\ data (black symbols)
with a power-law component partially covered by a distribution of neutral matter 
($\Gamma=1.61\pm 0.01$, $\nhsym= (2.5\pm0.1)\times 10^{22}\,$\nh, covering factor $CF=0.90\pm 0.01$). 

By fixing the parameters to the values found for \uno, we applied this simple model also to \due\ (red symbols in Fig.~\ref{fig:ra}, right panel) 
and to \tre\ (blue symbols).
The shape of the residuals are similar in all the three observations, confirming that the main driver of the observed variability 
is a change of the continuum intensity (by a factor of $\sim 1.5-1.7$).
Moreover, Fig.~\ref{fig:ra} clearly highlights that different and/or additional components are required to the fit
in order to proper reproduce the global spectral properties of the source during all the monitoring.
In particular, a line-like excess around $6.4\,$keV and a high-energy hump between $10$ and $30\,$keV are
visible, suggesting the presence of a reflection component.
In the softer part of the spectra (below $\sim 3\,$keV) the presence of different absorption troughs
could be indicative that we are intercepting an ionized material along the line of sight, possibly varying during the observations.
The presence of ionized materials, previously suggested by the \swift-XRT data \citep{severgnini15}, 
is also supported by the RGS results (see Section~\ref{sect:rgs}).

   \begin{figure}
   \centering
    \resizebox{1.\hsize}{!}{\includegraphics{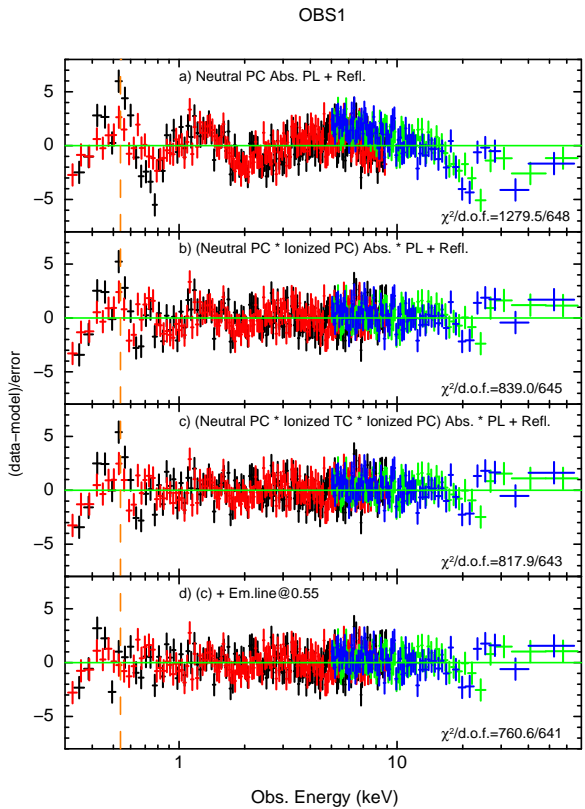}}
   \caption{Relevant residuals, plotted in terms of sigmas with error bars of size one, when different models are applied to the \xmm\ (black, pn; red, MOS) 
   and \nustar\ (green, FPMA; blue, FPMB) spectra of \src\ as recorded during the \uno\ observation: 
   from top to bottom, a continuum composed by the 
   intrinsic power law and a reflection component is view through a partially covering neutral absorber (panel a); in panel b) we present the residuals when
   a partially covering ionized absorber is included, while in panel c) a second ionized layer has been added.
   Finally, panel d) shows the residuals when the model includes also the soft X-ray emission line; its position is marked in all panels with a 
   vertical orange dashed line.
   All data are rebinned for clarity.
   }
              \label{fig:obs1_de}%
    \end{figure}

We first focused on 
the \uno\ spectrum at energies $>5\,$keV.
The addition of a Gaussian line to the absorbed power law results in a large improvement in the fit ($\dchidof=262.6/3$ between $5$ and $10\,$keV).
The profile of the \feka\ line [$E=6.42\pm 0.02\,$keV; $\sigma=86\errUD{25}{24}\,$eV; $F=(1.3\pm 0.2)\times 10^{-5}\,$\normGAUSS; 
EW$=125\errUD{31}{25}\,$eV] is marginally resolved.
Some residuals at $\sim 7.05\,$keV, the energy where the \fekb\ line is expected, can be accounted for by adding one more Gaussian line:
 $E=7.0\pm 0.1\,$keV; $\sigma$ tied to the width of the \feka\ line; EW$=21\errUD{15}{17}\,$eV; flux $\sim 14.8\,$per cent of the \feka\
\citep[in good agreement with the expectations;][]{palmeri03,yaqoob10}, although the improvement in the fit 
is not statistically significant ($\dchidof=4.5/2$ between $5$ and $10\,$keV).

The strength of the \feka\ line and the possible presence of a \fekb\ component 
support the presence of neutral reflection \citep{reynolds94,matt96,matt00}.
A contribution from a distant reflector was then included in the model by replacing the narrow Gaussian lines with a {\sc pexmon} model in {\sc xspec}  
\citep{pexmon}, an additive model self-consistently incorporating 
the Compton-reflected continuum from a neutral slab combined with emission from \feka, \fekb, \nika\ and the \feka\ Compton shoulder.
During the fits, we tied the {\sc pexmon} photon index and normalization to that of the primary power law, and we fixed the cutoff energy at $100\,$keV, 
the abundances of heavy elements at their Solar values, and
the inclination angle at $55\arcdeg$ \citep[see also \citealt{keel96,munoz07}]{severgnini15}.
The only free parameter of the {\sc pexmon} component was the reflection scaling factor,
found to be $\mathcal{R}=0.62\pm 0.08$ ($\pedap{\chi}{r}{2}=1.151$).

There is a small disagreement in photon index  ($\Delta \Gamma\sim 0.07$) between the \xmm\ and \nustar\
instruments, 
as previously reported for other simultaneous observations \citep[\eg,][]{parker16,cappi16}.
In order to take into account the remaining calibration uncertainties, in the following we allowed to vary 
the photon index between the satellites.
Here and below, we report $\Gamma$ and fluxes obtained for the \xmm\ only.

We then extended the analysis to the whole energy interval covered by the data, where we still have a 
poor fit below $\sim 3-4\,$keV ($\chidof=1279.5/648$; see Fig.~\ref{fig:obs1_de}, panel a).
Line-like structures, both in emission and in absorption, typically observed in partially/fully ionized absorber(s) flowing along the light of sight, 
could explain the observed residuals in the soft energy range. 
Absorption features at $E\sim0.8\,$keV associated with the presence of warm absorbers have been already detected in the \swift-XRT spectra of \src\ 
\citep{severgnini15}.
Since the presence of these components is also supported by the RGS data, we added a ionized absorber.
To model the absorption in \src\ in a physically consistent way, we used a grid of photoionized absorbers 
generated with the XSTAR\footnote{http://heasarc.gsfc.nasa.gov/docs/software/xstar/xstar.html} photoionization code
\citep{kallman04} assuming turbulence velocity of $\sigma=100\,$km~s$^{-1}$. 
This latter value is consistent with the narrow, or unresolved, widths of the absorption
lines detected in the RGS spectra (see Section~\ref{sect:rgs}).
The adopted XSTAR table assumes a power-law continuum ($\Gamma=2$) from $0.0136$ to $13.6\,$keV ionizing
a constant density shell of gas with standard solar abundances.
The free parameters of this component are the absorber column
density \nhsym\ and the ionization parameter\footnote{The ionisation parameter is defined as
$\xi=\pedix{L}{ion}/nR^2$, where $n$ is the hydrogen number density of the gas (in cm$^{-3}$) 
and $R$ is the radial distance of the absorbing/emitting material from the central source of X-ray (in cm), while
\pedix{L}{ion} is the ionising luminosity (in \lum) integrated between $1\,$Ryd and $1000\,$Ryd ($1\,\mbox{Ryd} = 13.6\,$eV).} $\log \xi/(\ionpar)$, 
varying in the ranges $10^{20}-10^{24}\,$\nh\ and $0-6$, respectively.

   \begin{figure}
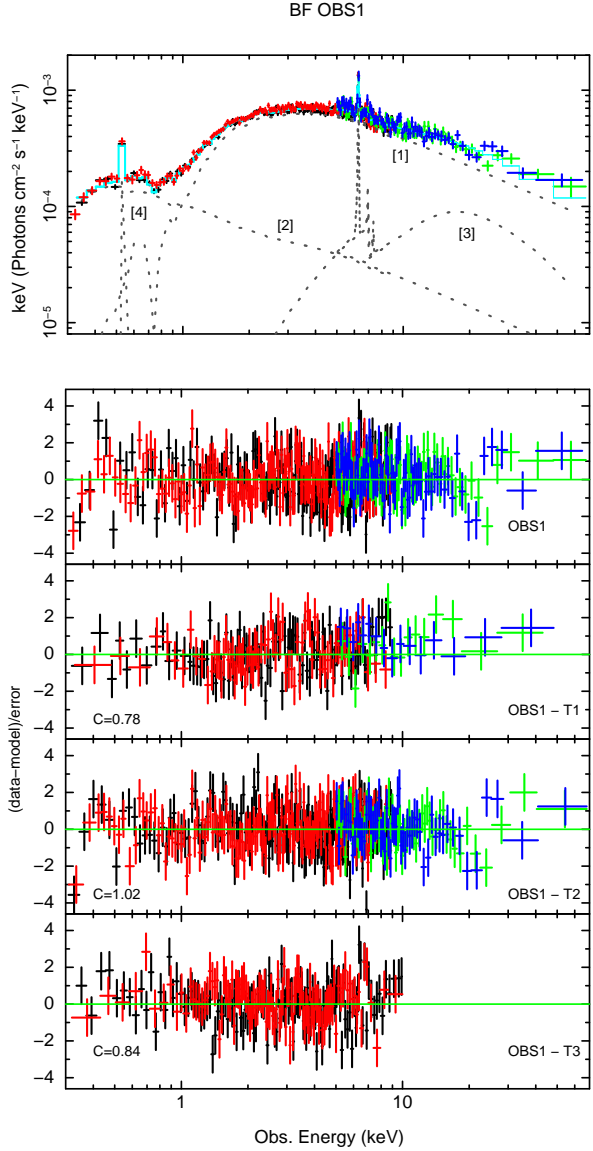

   \centering
    \resizebox{1.\hsize}{!}{\includegraphics{f6a.ps}}
    \resizebox{1.\hsize}{!}{\includegraphics{f6b.ps}}
   \vspace{0.1cm}
   \caption{Upper panel: unfolded \xmm\ (black, pn; red, MOS) and \nustar\ (green, FPMA; blue, FPMB) spectra of \src\ as recorded during the 
   whole \uno\ period,
   with the best-fit model applied (cyan continuous line): intrinsic power law partially covered by a warm absorber ([1] and [2] identify the fraction of continuum 
   covered and uncovered, respectively) seen through
   a second ionized layer of material, plus reflection from distant material [3] and narrow emission line [4]; the whole emission is absorbed by neutral material
   partially covering the central region.
   Lower panels, from top to bottom: relevant residuals, plotted in terms of sigmas with error bars of size one, when the model is applied to the 
   spectra observed 
   during the whole \uno\ period (same as in Fig.~\ref{fig:obs1_de}, panel d), 
   and to the three time intervals marked in Fig.~\ref{fig:lc} (as labelled), assuming a change in the intensity of the intrinsic continuum of a 
   factor $\mathcal{C}$ (with respect to the mean spectrum).
   }
              \label{fig:obs1}%
    \end{figure}

The results of the fit with two partially covering absorbers applied are a low-ionization state, $\log\xi/(\ionpar) = 0.71\pm 0.01$, 
with $\nhsym=(1.97\pm 0.02)\times 10^{22}\,$\nh, and a covering fraction\footnote{We define the covering fracion as the complement to one of the ratio 
between the normalizations of the unabsorbed and absorbed power laws.} $\simeq 0.94$;
and a higher column density 
[$\nhsym=(1.583\errUD{0.120}{0.127})\times 10^{23}\,$\nh] neutral absorber covering $35\errUD{3}{2}$\% of the central source.
The improvement in the fit is significant ($\chidof=840.0/645$); however, residuals below $\sim 2\,$keV are still visible (Fig.~\ref{fig:obs1_de}, panel b).
We thus tested the presence of one more layer of matter. 
This additional absorber, slightly more ionized [$\log\xi/(\ionpar) = 2.39\errUD{0.14}{0.09}$] has a lower column density 
[$\nhsym=(1.9\errUD{1.1}{0.7}) \times 10^{21}\,$\nh], and a covering fraction consistent with the unity; 
therefore, in the following analysis, we assumed that this absorber fully covers the whole emission.
The addition of the second layer improves the fit by $\dchidof=21.1/2$ (Fig.~\ref{fig:obs1_de}, panel c; see in particular at $E<2\,$keV).
We note that the presence of two layers of matter with different ionization states is suggested also by the high energy-resolution RGS data 
(see Section~\ref{sect:rgs}).
During the fit, both ionized absorbers have been assumed to be steady; allowing their velocities to vary, we found in both cases $\pedix{v}{out}<1000\,$km/s,
with no significant improvement in the fit.
Finally, 
residuals below $\sim1\,$keV require the addition of an emission line (parametrized with a narrow Gaussian, $\sigma$ fixed to $5\,$eV) 
to be accounted for.
The improvement in the fit is
significant ($\dchidof=57.3/2$; see Fig.~\ref{fig:obs1_de}, panel d); 
the energy is $E = 0.55 \pm 0.01\,$keV, with EW$=39\errUD{14}{16}\,$eV and $F=(2.9\errUD{0.8}{0.9})\times 10^{-3}\,$\normGAUSS.

In Fig.~\ref{fig:obs1} (upper panel) we have plotted the unfolded best fit to
the \xmm\ and \nustar\ data obtained with this combination of neutral and ionized absorbers and emission line, 
with the addition of reflection from distant matter; 
in the second panel we show the residuals between the data and the best-fitting model.
This complex model resulted in a roughly acceptable description of the data ($\chidof=760.6/641$); 
the best fit parameters are shown in Table~\ref{tab:bfpar}.

%
\begin{table*}
\begin{minipage}[t]{1\textwidth}
\caption{Analysis of the broad band \xmm-EPIC ($0.3-10\,$keV) and \nustar\ ($4-70\,$keV) spectra: result for a partial covering neutral 
absorber$*$\{ionized absorber$*$[partial covering ionized absorber$*$(intrinsic power law)]$+$reflection from  distant matter$+$narrow line\} model.}
\label{tab:bfpar}
\begin{center}
{
\footnotesize
\begin{tabular}{@{\extracolsep{0cm}}c@{\extracolsep{0.4cm}} c@{\extracolsep{0.13cm}} c@{\extracolsep{0.13cm}} c@{\extracolsep{0.18cm}} c@{\extracolsep{0.13cm}} c@{\extracolsep{0.18cm}} c@{\extracolsep{0.13cm}} c@{\extracolsep{0.13cm}} c@{\extracolsep{0.13cm}} c}
 \hline \hline
   & \multicolumn{3}{c}{\sc Direct and Reflected Continua} & \multicolumn{3}{c}{\sc Neutral and Ionized Absorbers} & \multicolumn{2}{c}{\sc Gaussian} & \\
  \cline{2-4} \cline{5-7} \cline{8-9}
   Obs.
   & $\Gamma$
   & Norm
   & $\mathcal{R}$
   & \nhsym
   & $\log\xi$
   & Cov.~fraction
   & \pedSM{E}{rf}
   & EW
   & \chidof \\
   (1)
   & (2)
   & (3)
   & (4)
   & (5)
   & (6) 
   & (7)
   & (8)
   & (9)
   & (10) \\
\vspace{-0.2cm}\\
\hline
\vspace{-0.2cm}\\
   \uno\ & $1.86\pm 0.03$ & $2.89\errUD{0.22}{0.16}$ & $0.68\pm 0.08$ & $14.37\errUD{2.94}{1.51}$ & $-$ 
   & $0.33\errUD{0.02}{0.03}$ & $0.55\pm 0.01$ & $39\errUD{14}{16}$ & $760.6/641$ \\
   &  &  &  & $0.13\errUD{0.14}{0.08}$ & $2.31\errUD{0.28}{0.40}$ & $1^{f}$ &  &  & \\
   &  &  &  & $1.92\errUD{0.05}{0.03}$ & $0.65\errUD{0.01}{0.09}$ & $0.94\pm 0.01$ &  &  & \\
\vspace{-0.2cm}\\
\hline
\vspace{-0.2cm}\\
   \due\ & $1.82\errUD{0.03}{0.02}$ & $1.88\errUD{0.08}{0.03}$ & $0.60\errUD{0.05}{0.07}$ & $14.37\errUD{3.20}{3.46}$ & $-$ 
   & $0.33\pm 0.02$ & $0.56\pm 0.01$ & $49\errUD{36}{14}$ & $658.6/553$ \\
   &  &  &  & $0.03\errUD{0.10}{0.01}$ & $<2.34$ & $1^{f}$ &  &  & \\
   &  &  &  & $1.67\pm 0.03$ & $0.43\errUD{0.02}{0.13}$ & $0.92\pm 0.01$ &  &  & \\
\vspace{-0.2cm}\\
\hline
\vspace{-0.2cm}\\
   \tre\ & $1.85\pm 0.01$ & $1.86\errUD{0.04}{0.03}$ & $0.54\errUD{0.13}{0.10}$ & $15.28\errUD{4.80}{4.38}$ & $-$ 
   & $0.31\errUD{0.03}{0.05}$ & $0.56\errUD{0.05}{0.09}$ & $<40$ & $479.0/463$ \\
   &  &  &  & $<5.25$ & $<3.14$ & $1^{f}$ &  &  & \\
   &  &  &  & $2.33\errUD{0.09}{0.06}$ & $0.59\errUD{0.15}{0.08}$ & $0.92\pm 0.01$ &  &  & \\
\end{tabular}
 }
 \end{center}       
 {\footnotesize   {\sc Note:} Errors are quoted at the 90\% confidence level for 1 parameter of interest ($\Delta\apix{\chi}{2}=2.71$). $f=\:$parameter fixed 
 during the fit. The neutral Compton reflection, described using the {\sc pexmon} model in {\sc xspec} include, self-consistently both the continuum and 
 the Fe and Ni emission lines.
 \footnotesize Column (1): observation.
 \footnotesize Column (2): intrinsic power law and reflection component photon index.
 \footnotesize Column (3): intrinsic power-law normalization, in units of $10^{-3}\,$\normPL\,@$1\,$keV.
 \footnotesize Column (4): reflection fraction, with respect to the intrinsic continuum observed during \uno.
 \footnotesize Column (5): warm absorber column density, in units of $10^{22}\,$\nh.  
 \footnotesize Column (6): ionization parameter $\xi=\pedix{L}{ion}/nR^2$ (in units of \ionpar), where \pedix{L}{ion} is 
 the ionising luminosity (in units of \lum), $n$ 
 is the hydrogen number density of the gas (part~cm$^{-3}$) of the illuminated slab, and $R$ is the radial distance of the absorbing/emitting material 
 from the 
 central source of X-ray (in cm).
\footnotesize Column (7): complement to one of the uncovered-to-covered flux ratio.
 \footnotesize Column (8): rest-frame energy centroid of the Gaussian emission line (width fixed to $50\,$eV), in units of keV.
 \footnotesize Column (9): emission line equivalent width, in units of eV.
 \footnotesize Column (10): $\chi^2$ and number of degrees of freedom.
 }
\end{minipage}
\end{table*}

   \begin{figure}
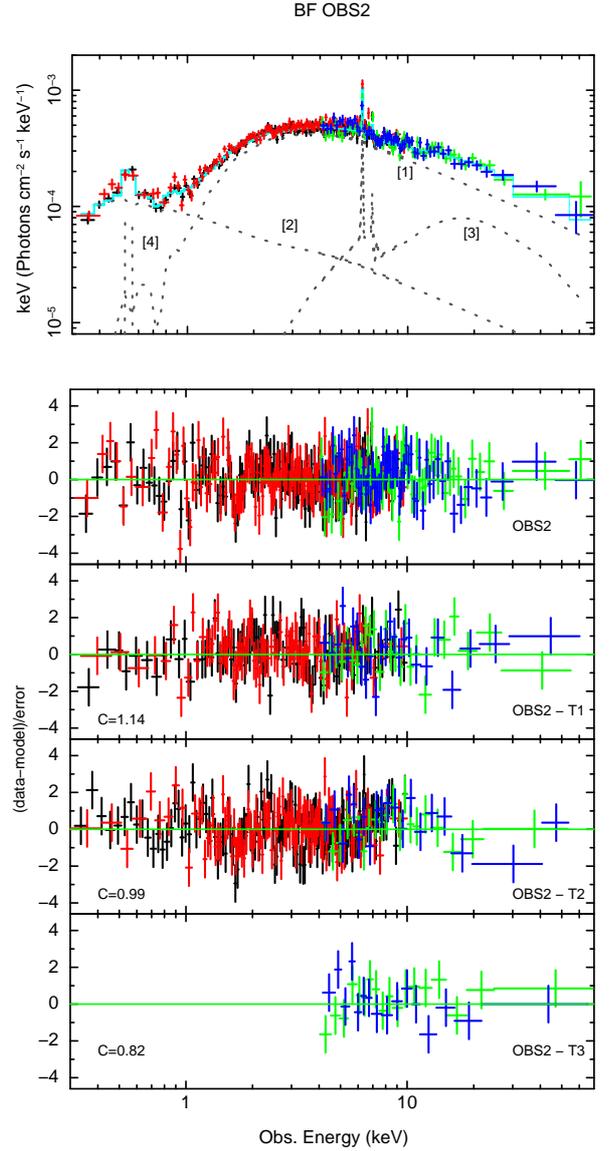

   \centering
    \resizebox{1.\hsize}{!}{\includegraphics{f7a.ps}}
    \resizebox{1.\hsize}{!}{\includegraphics{f7b.ps}}
   \caption{As in Fig.~\ref{fig:obs1}, but for the \due\ period.}
              \label{fig:obs2}%
    \end{figure}
%

   \begin{figure}
   \centering
    \resizebox{1.\hsize}{!}{\includegraphics{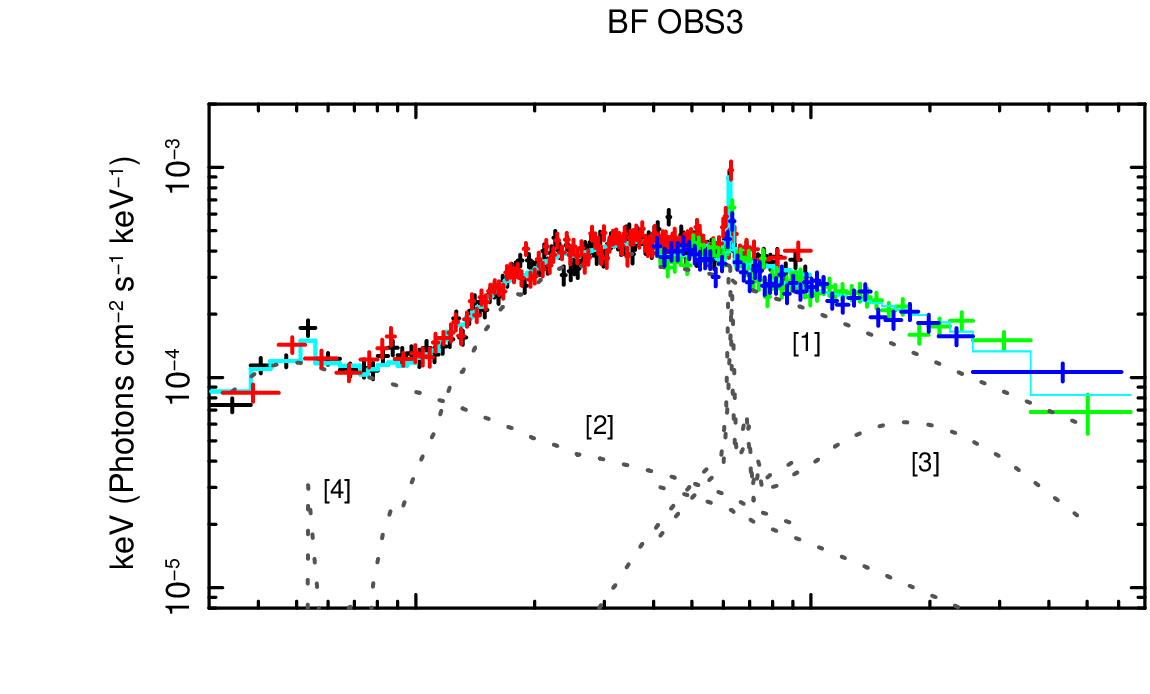}}
    \resizebox{1.\hsize}{!}{\includegraphics{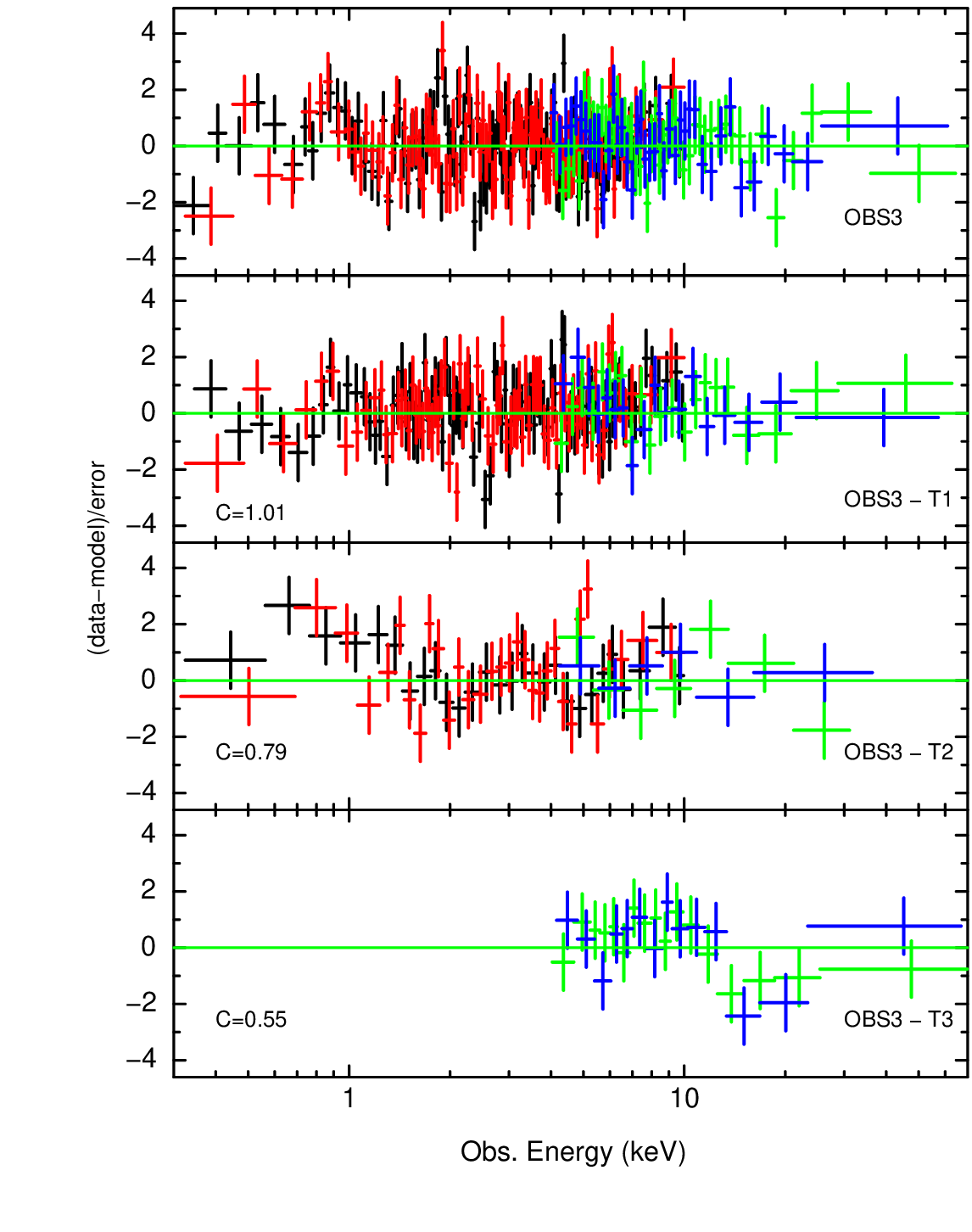}}
   \caption{As in Fig.~\ref{fig:obs1}, but for the \tre\ period.}
              \label{fig:obs3}%
    \end{figure}
%

%
\begin{table}
\begin{minipage}[t]{0.5\textwidth}
\caption{X-ray fluxes and luminosities.}
\label{tab:lumfl}
 \begin{center}
{
\tiny
\begin{tabular}{@{\extracolsep{0cm}}c@{\extracolsep{0.1cm}} c@{\extracolsep{0.08cm}} c@{\extracolsep{0.08cm}} c@{\extracolsep{0.15cm}} c@{\extracolsep{0.08cm}} c@{\extracolsep{0.08cm}} c@{\extracolsep{0.08cm}} c@{\extracolsep{0cm}}}
  \hline \hline
   &  \multicolumn{3}{c}{\sc Observed Flux} & \multicolumn{4}{c}{\sc Intrinsic Luminosity} \\     
\vspace{-0.15cm}\\
    \cline{2-4}\cline{5-8}          
\vspace{-0.15cm}\\
    Obs.
    & {\tiny 0.3-2 keV}
    & {\tiny 2-10 keV}
    & {\tiny 10-70 keV}
    & {\tiny 0.3-2 keV}
    & {\tiny 2-10 keV}
    & {\tiny 10-70 keV}
    & {\tiny 0.0136-13.6 keV}\\
\vspace{-0.15cm}\\
    \cline{1-8}          
\vspace{-0.15cm}\\
    \uno\ & $0.96\errUD{0.11}{0.06}$ & $7.66\pm 0.60$ & $19.85\errUD{1.86}{1.90}$ & 
    $10.53\errUD{0.63}{0.74}$ & $11.21\errUD{0.67}{0.79}$ & $18.53\errUD{1.11}{1.31}$ & 36.88$\errUD{2.20}{2.60}$ \\
\vspace{-0.15cm}\\
    \due\ & $0.64\errUD{0.02}{0.08}$ & $5.49\errUD{0.23}{0.25}$ & $13.70\errUD{0.07}{0.08}$ & 
    $6.66\errUD{0.27}{0.09}$ & $7.68\errUD{0.31}{0.11}$ & $11.67\errUD{0.47}{0.16}$ & 23.30$\errUD{0.94}{0.32}$ \\
\vspace{-0.15cm}\\
    \tre\ & $0.56\pm 0.02$ & $5.02\errUD{0.22}{0.17}$ & $13.56\errUD{1.10}{0.85}$ & 
    $6.64\errUD{0.14}{0.10}$ & $7.34\errUD{0.16}{0.11}$ & $11.88\errUD{0.25}{0.18}$ & 23.21$\errUD{0.49}{0.35}$ \\
\end{tabular}
}
 \end{center}       
 {\footnotesize   {\sc Note:} Errors are quoted at the 90\% confidence level for 1 parameter of interest ($\Delta\apix{\chi}{2}=2.71$). 
 Luminosity are absorption-corrected, while fluxes are corrected only for the Galactic absorption. 
 Fluxes (in units of $10^{-12}\,$\flux) and luminosities (in units of $10^{42}\,$\lum) refer to the pn.
 }
\end{minipage}
\end{table}
%

\subsection{The origin of the X-ray variability: comparison among the different datasets}\label{sect:all}

Having found a physical scenario able to describe the \uno\ data, we explored the possibility of explaining the variability observed by changing only
the best fit parameters (but assuming the same components in the model): therefore, we applied the model described in Sect.~\ref{sect:bb} 
to the \due\ and \tre\ datasets.

As a first step, we fitted the data twice, by leaving free to vary the intrinsic power-law normalization only or the warm absorbers parameters only,
respectively.
We found that the observed variations can be attributed mainly to a decreasing of a factor $\sim 1.5-1.7$ 
of the direct continuum, although residuals observed below $\sim 2\,$keV are indicative of some variations occurred in the absorbing medium(s). 

This is confirmed also by analysing the \due\ and \tre\ states leaving free to vary all the different components of the model.
Note that, to better compare the reflection fractions in the three observations, both \due\ and \tre\ data have been fitted 
by fixing the normalization of the {\sc pexmon} model to the value of \uno.
The best fit value for the parameters are reported in Table~\ref{tab:bfpar}, while Fig.~\ref{fig:obs2} and Fig.~\ref{fig:obs3} show the unfolded spectra
(upper panel) and the residuals (second panel) for \due\ ($\chidof=658.6/553$) and \tre\ ($\chidof=479.3/463$), respectively.
Table~\ref{tab:lumfl} presents the observed fluxes (corrected for Galactic absorption) and unabsorbed luminosities for the three observations.

As discussed in Sect.~\ref{sect:lc}, small but statistically significant variations are observed in different energy ranges
during each observation, although with substantially constant HRs. 
For each observation, we applied the model discussed so far to the 
datasets collected in the T1, T2, and T3 intervals, as marked in Fig.~\ref{fig:lc}.
By fitting the data jointly, we tried to reproduce the data leaving free to vary between the single intervals 
the intrinsic power-law normalization only or the absorbers parameters only.
We found that a variation in the intrinsic continuum is always required,
confirming that 
also during each observation
a change in the continuum intensity 
of a factor $\mathcal{C}$ (with respect to the mean spectrum) ranging from $\sim 1.14$ to $\sim 0.55$
is the main driver of the light-curve properties.
The relevant residuals are plotted in the last three panels of Fig.~\ref{fig:obs1} (\uno), Fig.~\ref{fig:obs2} (\due) and Fig.~\ref{fig:obs3} (\tre).

Finally, we tested the EPIC$+$\nustar\ best-fit model with the RGS spectra.
The data do not allow us to properly fit such a complex model, 
therefore we 
allowed only one absorbing layer to vary at each time, keeping
the remaining components 
fixed
to the best fit values reported in Table~\ref{tab:bfpar}.
We found ionization parameters and column densities broadly consistent with the values previously derived.
For the partially-covering absorber we found 
$\log \xi/(\ionpar)=0.63\errUD{0.04}{0.08}$, $\nhsym=(2.0\pm0.1)\times10^{22}\,$\nh; 
$\log \xi/(\ionpar)=0.6\errUD{0.1}{0.2}$, $\nhsym=(2.1\errUD{0.4}{0.2})\times10^{22}\,$\nh;
$\log \xi/(\ionpar)<0.5$, $\nhsym=(1.5\errUD{0.6}{0.3})\times10^{22}\,$\nh, for \uno, \due\ and \tre, respectively.
While for the total covering absorber we obtained: 
$\log \xi/(\ionpar)<1.7$, $\nhsym<5\times10^{21}\,$\nh; 
$\log \xi/(\ionpar)<1$, $\nhsym<3\times10^{21}\,$\nh; 
$\log \xi/(\ionpar)<1$, $\nhsym<4\times10^{21}\,$\nh, for \uno, \due\ and \tre, respectively.

\section{Discussion}\label{sect:disc}

We have presented a detailed spectral analysis of the \xmm\ $+$ \nustar\ monitoring of \src, covering slightly more than $11\,$days.

\begin{figure}
   \centering
    \resizebox{1.\hsize}{!}{\includegraphics[angle=270]{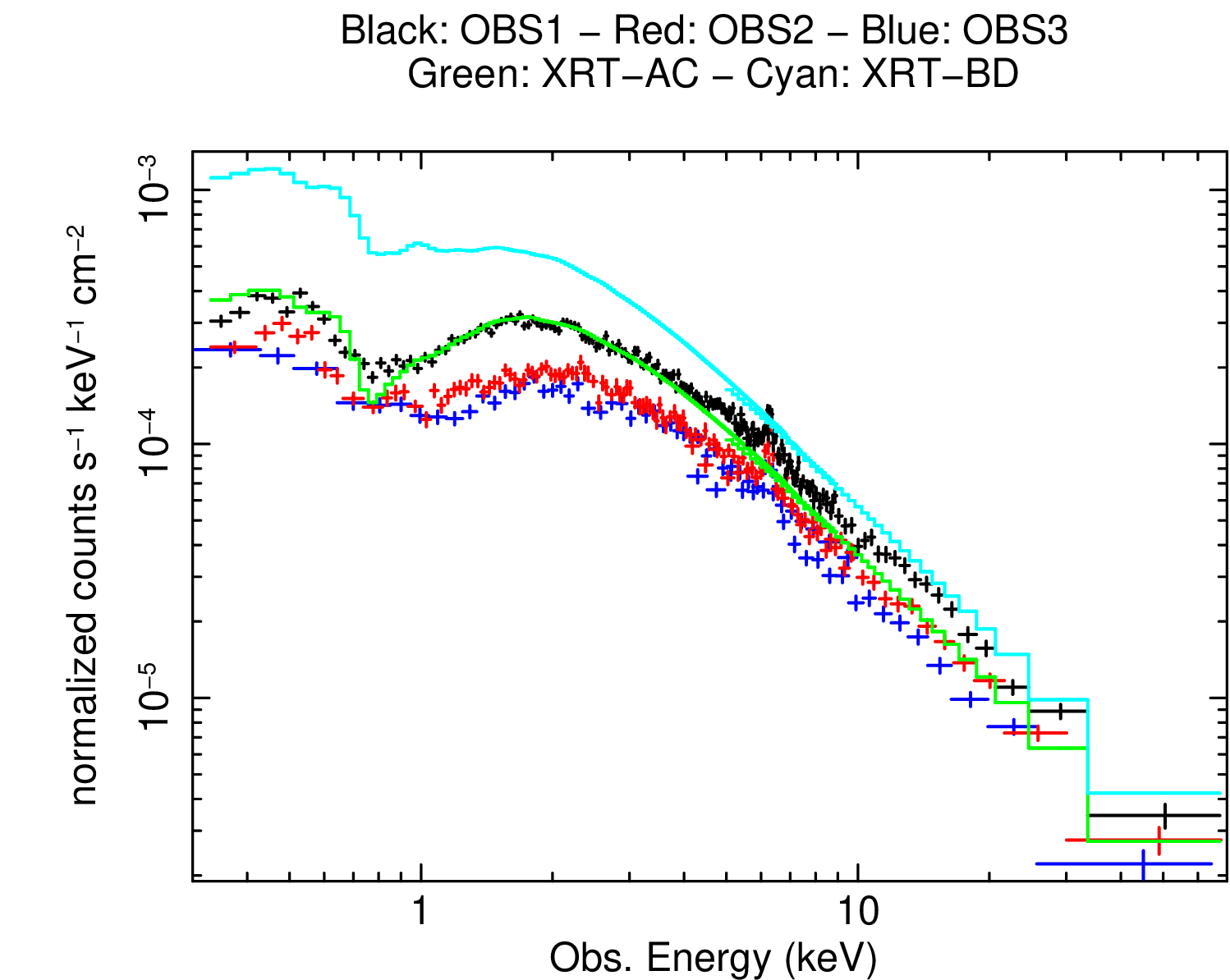}}
   \caption{\xmm\ and \nustar\ spectra (black, \uno; red, \due; blue, \tre) compared with the two emission states observed during the \swift-XRT 
   monitoring, as analysed in \citet[green, low state; cyan, high state]{severgnini15}. 
   }
              \label{fig:cfr}%
    \end{figure}

Several features are clearly evident in the low energy resolution spectra of \src\ (see Fig.~\ref{fig:ra}): a strong absorption 
between $\sim 0.7$ and $\sim3\,$keV, 
a spectral flattening at lower energies, emission-line residuals in the iron band, and a bump at energies higher than $\sim 10\,$keV.
Moreover, variations in intensity in the soft, medium, and hard energy ranges, but not in spectral shape are clearly seen (see Fig.~\ref{fig:lc}).
However, the level of variation is much lower than observed in the previous \swift-XRT monitoring \citep{severgnini15}: in Fig.~\ref{fig:cfr} 
we compare our new data with the models 
which provide the best fit to
the two states identified in the XRT data.
As evident, the new monitoring finds the source at an emission level comparable with the low state observed by XRT, 
while the high XRT state is never reached.

The analysis of the RGS data provides an hint for the presence of a multi-layer partially ionized absorber, 
as previously suggested by \citet{severgnini15}; a few emission lines are also detected.
The $0.3-70\,$keV \xmm-EPIC and \nustar\ spectra have been analyzed 
in the framework of an ``absorption-based'' scenario.
In our best-fit broad-band model, the primary power-law emission is partially covered (covering fraction $\sim 0.9$) by a 
low ionization ($\xi\sim 3-5\,$\ionpar) warm absorber, and totally covered by a 
mildly-ionized ($\xi\lesssim 200\,$\ionpar) absorber.
A relatively small fraction (covering factor $\sim 0.3$) of the whole emission is 
absorbed by a rather high ($\sim 1.5\times 10^{23}\,$\nh) 
column density of neutral gas.
A narrow Gaussian emission line, consistent with K$\alpha$ fluorescence from almost neutral iron, accounts for the residuals observed 
between $6$ and $7\,$keV,
while the reflection component expected to be associated to the line can reproduce the spectral curvature observed at high energy.
Finally, a narrow emission line is detected at $\sim 0.56\,$keV; the most likely identification of this feature is the \ion{O}{vii}\,\hea.
The presence of this line is supported also by the RGS data.

\begin{figure*}
   \centering
    \resizebox{1.\hsize}{!}{\includegraphics{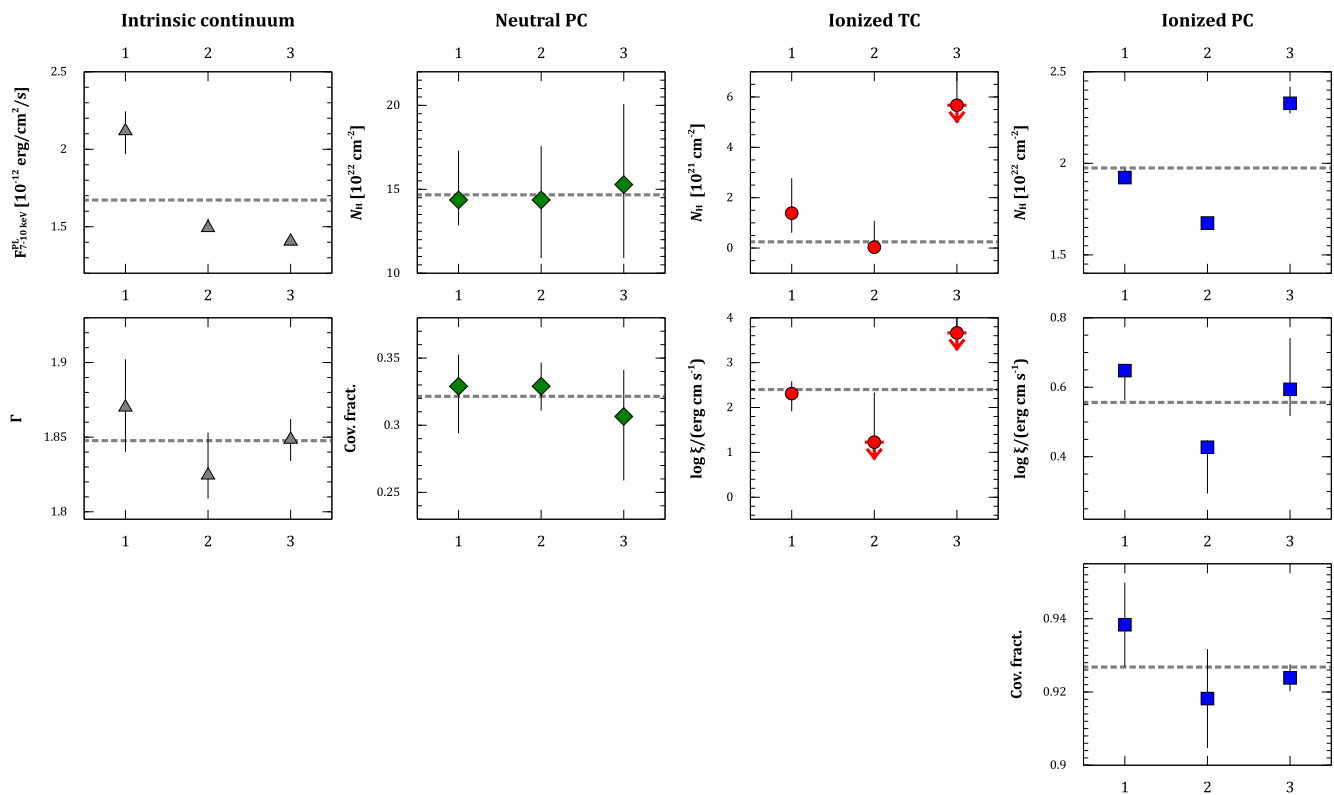}}
   \caption{
   Evolution of spectral fit parameters versus the observation number. 
   From left to right: 
   intrinsic properties (grey triangles): power-law flux between $7$ and $10\,$keV (in units of $10^{-12}\,$\flux) and $\Gamma$;
   partial covering neutral absorber (green diamonds): \nhsym (in units of $10^{22}\,$\nh) and covering fraction; 
   fully covering warm absorber (red circles): \nhsym (in units of $10^{21}\,$\nh), $\log \xi$; 
   partial covering warm absorber (blue squares): \nhsym (in units of $10^{22}\,$\nh), $\log \xi$, covering fraction. 
   Dashed lines mark the mean value of each parameter.}
              \label{fig:var}%
    \end{figure*}

Given the complexity of the multi-component spectral model used here, we do not expect that the observed spectral variability could be ascribed 
to a single parameter alone, but rather to several parameters combined.
In Fig.~\ref{fig:var} we plot the evolution of the spectral fit parameters.
Passing from \uno\ to \due\ and \tre\ observations, the direct continuum decreased significantly, as evident from the top-left panel where we plot the intrinsic 
$7-10\,$keV 
flux light curve: the flux in this energy band should have very little, if any, sensitivity to the obscuration.
The strength of the reflection component is basically constant:
the reflection fraction is consistent between the three periods
($\pedix{\mathcal{R}}{\uno}=0.68\pm 0.08$, $\pedix{\mathcal{R}}{\due}=0.60\errUD{0.05}{0.07}$, and 
$\pedix{\mathcal{R}}{\tre}=0.54\errUD{0.13}{0.10}$, see Table~\ref{tab:bfpar}), 
implying that this component does not respond to the variation of the direct emission,
in agreement with an origin of the reflection in rather distant matter.

In addition, slight variations in the column density and ionization state of the partially-covering absorber are observed (right column in Fig.~\ref{fig:var}). 
As regards the second ionized absorber, instead, the data are consistent with a substantially constant component, although poorly constrained 
in \due\ and \tre\
(third column in Fig.~\ref{fig:var}).
Finally, as evident in the second column of Fig.~\ref{fig:var}, the neutral absorber does not change between the observations.

Despite being a clear over-parametrization of the \swift-XRT data, the same model can be used to satisfactorily reproduce also the two \swift-XRT states.
We fixed the components of the model that, from our analysis of the \xmm$+$\nustar\ monitoring, are consistent with being constant: 
the reflection component, the neutral absorber, and the total covering ionized absorber.
With these assumptions, the change from the low to the high state can be explained by a concomitant variation 
in the intrinsic flux of a factor $\sim 1.6$ and a change of the column density of the partially covering low-ionized absorber 
($\Delta\nhsym\sim5\times10^{21}\,$\nh).

In the following we will try to set some plausible constraints on
the location $R$ of the two 
ionized absorbers
as a function of the density of the medium $n$ 
using the relation with the ionization luminosity and the ionization parameter, $\xi=\pedix{L}{ion}/nR^2$.
As for
the total covering  mildly-ionized absorber
we focus on \uno, 
since this is
the only observation in which the parameters of 
this component are well-constrained (see Table~\ref{tab:bfpar}).
We estimated a
ionizing luminosity between $13.6\,$eV and $13.6\,$keV, $\pedix{L}{ion,\uno}\sim 3.7\times 10^{43}\,$\lum\ 
(see Table~\ref{tab:lumfl}), thus implying:
\begin{equation}\label{eq:tc} 
nR^2=2\times 10^{41}\,\mbox{cm}^{-1}
\end{equation}
As for 
the partially-covering low-ionized layer,
from the best-fit values found for each observation as reported in Table~\ref{tab:bfpar}, 
and given the corresponding ionizing luminosities between $13.6\,$eV and $13.6\,$keV 
observed during the monitoring, $\pedix{L}{ion}\sim (2.3-3.7)\times 10^{43}\,$\lum\ (see Table~\ref{tab:lumfl}), we have: 
\begin{equation}\label{eq:pc}
nR^2=(6-9)\times 10^{42}\,\mbox{cm}^{-1}
\end{equation}
By considering  typical density values  for the  BLR (\pedix{n}{BLR}), the torus (\pedix{n}{torus}) and 
the NLR (\pedix{n}{NLR}), we checked if the values of $R$ derived by equations~\ref{eq:tc} and~\ref{eq:pc} are consistent
with the expected distances of these regions from the central AGN in \src.
For a typical BLR density $\pedix{n}{BLR}> 10^8\,$cm$^{-3}$ \citep{osterbrock89,netzer13} 
we obtain $R\lesssim 4\times 10^{16}\,$cm and $R\lesssim 3\times 10^{17}\,$cm
for the total and the partial covering absorbers, respectively.
These values are
consistent with the distance of the BLR as estimated from the optical luminosity \citep[$\pedix{R}{BLR}\approx 10^{16}-10^{17}\,$cm;][]{severgnini15}.
Similarly, by assuming
$\pedix{n}{torus}\sim 10^6\,$cm$^{-3}$ \citep{netzer13}, the derived 
distances are
$R\sim 4\times 10^{17}\,$cm and $R\sim 3\times 10^{18}\,$cm 
for the total and the partial covering absorbers, respectively,
in good agreement with 
the inner radius of a torus as derived from observations in the infrared
(pc-scale, see \eg\ \citealt{jaffe04}, and the discussion in \citealt{burtscher13}).
Assuming instead the NLR density quoted by \citet{bennert06} for \src, $\pedix{n}{NLR}\sim 500-1000\,$cm$^{-3}$, 
we find $R\sim (1-2)\times 10^{19}\,$cm and $R\sim (0.8-1)\times 10^{20}\,$cm (total and partial covering absorbers, respectively).
These values are  about 
a factor 
$800-100$
lower than the radius of the NLR estimated by the same authors, $\pedix{R}{NLR}\sim 8\times 10^{21}\,$cm.

These considerations imply that the two ionized absorbers could be part of the 
BLR or they could be distributed between the BLR and the inner part of the 
neutral torus, well inside the radius of the NLR. 

Similar, but less stringent, results can be obtained 
under the reasonable assumption of the thickness of each layer $\Delta R\sim\nhsym/n$ lower than $R$.
In this case, we can derive an upper limit to the distance $R<\pedix{L}{ion}/(\nhsym \xi)$: we found
$R<1.4\times 10^{20}\,$cm and $R<5.2\times 10^{20}\,$cm for the totally-covering and the partially-covering absorbers, respectively, 
well within the NLR in \src.
Multi-layer absorbers, possibly stratified, characterized by different ionization states and densities, located at distances 
ranging from tens to hundreds of parsecs from the central source have been detected in several Seyfert galaxies
\citep[see \eg][]{ebrero10,ebrero13}.

As for the neutral absorber, the high column density, low covering factor and constancy of 
this component can suggest its association with the circumnuclear torus.
In this case, 
as suggested by \citet{severgnini15},
our line of sight likely grazes (and partly intercepts) the walls of the torus, 
providing us a favoured 
perspective to detect internal multi-zone ionized absorbers.

An alternative possibility 
is that all the different absorbers, neutral and ionized, here detected are part of the same obscuring medium: 
it would consist of a mix of gas with different temperature and density, as observed \eg\ in NCG~5548 \citep{kaastra14}.

Unfortunately, with the present data we are not in the position to discriminate between these different scenarios.

%

\section{Conclusions}\label{sect:concl}

We have performed a broad-band monitoring programme of \src, observed jointly with the \xmm\ and \nustar\ satellites.
The programme consists of $3$ observations separated from each other by about $5\,$days. 
The main results of this work can be summarized as follows.

   \begin{enumerate}
      \item This monitoring found \src\ in an emission level comparable with the low state observed by \swift-XRT \citep{severgnini15}, 
      while the high XRT state is never reached.
      Comparing the X-ray light curves in the soft ($0.3-1\,$keV, $1-2\,$keV, and $0.3-3\,$keV), medium ($3-10\,$keV) and hard ($10-70\,$keV) 
      energy bands, we found variations in intensity, but not in spectral shape.
      The strongest variation is observed between \uno\ and \due, when the count rate decreases by a factor of $\sim1.4$.
      \item The high energy resolution data collected by the RGS suggest the presence of a multi-layer partially ionized absorber.
      A few emission lines are also observed.
      \item As for the low-resolution 
      spectra, the importance of absorption structures is evident in all datasets.
      The three states can be well described within an ``absorption-based'' scenario, fully consistent with the results obtained from the RGS data: 
      an intrinsic power law with 
      photon index $\Gamma \sim 1.85$ affected by a two-phase warm absorber with slightly different ionization state and column density, 
      covering the central source in different ways; 
      a high-column density neutral absorber intercepts a small fraction of the the central emission. 
      The narrow \feka\ emission line and the excess in the continuum above $\sim 10\,$keV can be explained with a cold reflection from distant matter; 
      the strength of this component is constant during the monitoring.
     \item The main driver of the observed variations is a decreasing of the direct continuum; slight variations in the partial covering ionized 
     absorber are also detected.
     As regarding the total covering absorber, it is firmly detected only in the first observation; however, the data are consistent with no variation 
     of this component.
     The best-fit properties derived from our analysis suggest a location of the two ionized absorbers consistent both with the BLR, 
     the innermost part of the torus, or in between. 
     As regards the neutral component, almost constant during the monitoring, it 
     may either be part of the same stratified structure or associated with the walls of the torus.
     Based on the present data, we are not able to distinguish between the different hypotheses: a) a line of sight
     passing through different parts of a stratified structure of ionized gas with embedded colder and denser
     parts, or b) a line of sight partly intercepting the walls of the torus, providing us a favoured perspective to observe the
     structure of a multi-zone ionized absorber.
  \end{enumerate}
%

\section*{Acknowledgements}
     We are grateful to the referee for her/his constructive comments that improved the paper.
     We warmly thank A. Caccianiga for the useful discussions and the helpful suggestions.
     
     Support from the Italian Space Agency is acknowledged 
     (contract ASI INAF NuSTAR I/037/12/0).
      
      Based on observations obtained with \xmm, an ESA science mission with instruments and contributions directly funded by 
      ESA Member States and the USA, NASA.
      This work made use of data from the \nustar\ mission, a project led by the California Institute of Technology, managed by 
      the Jet Propulsion Laboratory, and
      funded by NASA. 
      This research has made use of the  \nustar\ Data Analysis Software ({\sc NuSTARDAS}) jointly developed by the ASI Science Data Center and 
      the California Institute of Technology.

     This research has made use of NASA's Astrophysics Data System.



\bibliographystyle{mnras} 
\bibliography{ms_mrk915_ref} 



%


\label{lastpage}
\end{document}